\documentclass{aa}  

\usepackage{graphicx}
\usepackage{txfonts}
\usepackage{lipsum}
\usepackage{subcaption}         
\usepackage{lscape}             
\usepackage{placeins}           
\usepackage[usenames,dvipsnames]{color} 

  \usepackage[
  breaklinks = true,
  colorlinks = true,
  urlcolor   = blue,
  citecolor  = blue,
  linkcolor  = blue,
]{hyperref}

\begin{document}

   \title{Formation and rising phase of a flux rope through data-constrained simulations}



   \author{M.~V.~Sieyra\inst{1}
        \and A.~Strugarek\inst{1}
        \and A.~Prasad\inst{2,3,4}
        \and A.~Wagner\inst{5}
        \and P.~D\'emoulin\inst{6}
        \and F.~Moreno-Insertis\inst{7,8}
        \and A.~J.~Finley\inst{1,9}
        \and R.~Joshi\inst{7,8,3,4}
        \and A. Blaise\inst{1}
        \and A.~S.~Brun\inst{1}
        \and E.~Buchlin\inst{10}
        }

   \institute{
   $^{1}$ D\'epartement d'Astrophysique/AIM, CEA/IRFU, CNRS/INSU, Universit\'e Paris-Saclay, Universit\'e de Paris Cit\'e, Orme des Merisiers, Bat. 709,  
   91191 Gif-sur-Yvette Cedex, France, \email{vsieyra@unc.edu.ar}\\ 
   $^{2}$  Statkraft AS, Lysaker, Norway \\
   $^{3}$ Institute of Theoretical Astrophysics, University of Oslo, P.O. Box 1029 Blindern, N-0315 Oslo, Norway\\
    $^{4}$ Rosseland Centre for Solar Physics, University of Oslo, P.O. Box 1029 Blindern, N-0315 Oslo, Norway \\
   $^{5}$ Department of Physics, University of Helsinki, Helsinki, Finland \\
   $^{6}$ LIRA, Observatoire de Paris, Université PSL, CNRS, Sorbonne Université, Université Paris-Cité, France\\
   $^{7}$ Instituto de Astrofisica de Canarias, E-38205 La Laguna, Tenerife, Spain\\
   $^{8}$ Departamento de Astrofisica, Universidad de La Laguna, E-38206 La Laguna, Tenerife, Spain\\
   $^{7}$ Heliophysics Science Division, NASA Goddard Space Flight Center, Greenbelt, MD 20771, USA \\
   $^{8}$ Department of Physics and Astronomy, George Mason University, Fairfax, VA 22030, USA \\
   $^{9}$ European Space Agency, ESTEC, Noordwijk, The Netherlands \\
   $^{10}$ Université Paris-Saclay, CNRS, Institut d’Astrophysique Spatiale, 91405 Orsay, France \\
            }

   \date{Received XX, 20XX}

 
  \abstract
   {Advances in data-constrained and data-driven simulations have shed light on the initiation of solar eruptions.
   These models incorporate observed photospheric magnetic fields. However, due to the lack of magnetic field information in the rest of the solar atmosphere, models rely on extrapolations that, in most cases, neglect the Lorentz force. Nevertheless, this force is present in the lower atmosphere and may play a key role in destabilizing the equilibrium configuration and triggering eruptions.}
   {This study seeks to understand and reproduce a solar eruption SOL2014-12-18T21:41 that occurred in active region NOAA 12241, 
   preceded by an M6.9 flare, and to investigate the impact of relaxing the initial force-free assumption.} 
   {The resistive and compressible magnetohydrodynamic simulation is initiated using a non-force-free magnetic field extrapolated from a photospheric vector magnetogram taken minutes before the flare. The simulation includes a stratified atmosphere and non-ideal effects such as thermal conduction and radiative cooling. }
   {A flux rope forms and rises in the simulation, carrying away dense material from the lower solar atmosphere. Its formation results from the non-zero Lorentz force acting on the initial sheared arcade, without assuming pre-existing flux ropes or photospheric driving motions. The flux rope is then deflected toward regions of low magnetic pressure, escaping the domain at 350~km~s$^{-1}$ with approximately constant acceleration.}
   {A robust numerical framework for modelling flaring active regions was applied to the eruption of NOAA AR12241 as a case study,
   assuming a realistic non-force-free magnetic field near the flare onset. It exemplifies how an initial Lorentz force imbalance can successfully trigger a flux rope formation that later escapes the simulation domain. It also enables comparison with real observations through the addition of a stratified atmosphere spanning from the photosphere to the corona.}


   \maketitle

\section{Introduction}
\label{sec:intro}

Solar eruptions are one of the most studied dynamical events in the solar atmosphere. These explosive events are usually associated with flares, coronal mass ejections (CMEs) and energetic particles.
During flares, magnetic energies between $10^{28}$ and $10^{31}$~erg are released and converted into kinetic, thermal and non-thermal energies \citep{Shibata2011,Fletcher2011,Benz2017}. As a consequence, they can lead to extreme ultraviolet and X-ray radiation, particle acceleration and material ejection (CMEs), all of which can dramatically affect space weather. 

There are two main categories of eruption initiation \citep{Jiang2024}: one is based on the ideal magnetohydrodynamic (MHD hereafter) instability of a pre-existing magnetic flux rope \citep{TD1999,Titov2018}, and the other based on reconnection of sheared field lines with or without a flux rope. 
Among the MHD instabilities, there are two commonly considered, the kink instability \citep{Hood&Priest1979, Torok2004, Ji2003} and the torus instability \citep{Kliem2006}. In their simplest form, these instabilities are triggered when the magnetic configuration exceeds a threshold, namely, concerning the total number of magnetic windings between the line-tied ends for the kink instability \citep{Hood&Priest1979} or the decay index with height of the external strapping magnetic field for the torus instability \citep{Kliem2006}.
Within the reconnection-based models, we have, roughly speaking, the tether-cutting \citep{Moore2001} and the breakout model \citep{Antiochos1999}. 
An example of the tether-cutting mechanism  is the standard flare model \citep[e.g.][]{Carmichael1964, Sturrock1965, Hirayama1974, Kopp&Pneuman1976, Janvier2014}, in which there is a pre-existing magnetic flux rope. 

Usually, in numerical simulations, we can find a combination of these ideal and reconnection processes together with triggering mechanisms at photospheric or chromospheric levels such as shearing motions \citep{Lynch2008, Pariat2010, Wyper2017, RJoshi2024}, flux cancellation \citep{Amari2003b, Linker2003, Aulanier2010, Zuccarello2015} and flux emergence \citep{Leake2010, Leake2014, Leake2022, MacTaggart&Haynes2014}. 

In real events the process is even more complex, and eruptions may not obey these idealised theoretical mechanisms. It remains unclear how solar eruptions are initiated, especially because of the lack of direct observational information about the magnetic field strength and orientation in the corona. Therefore, in order to better understand the mechanisms responsible for solar eruptions, data-based models constrained by photospheric observations have been shown to be an appropriate path. The readers are referred to review papers of \citet{Inoue2016,Jiang2022} for such data-based simulations. We can distinguish between data-constrained, where the initial conditions are derived from observational data at the solar surface at a single time, and data-driven simulations that use time-dependent boundary conditions derived from observational time series. 

Focusing on the eruption process, data-constrained simulations are generally initialised with close to unstable magnetic configurations extrapolated from vector magnetograms at a moment shortly before eruption onset. For example, using non-linear force-free extrapolations (NLFFF hereafter), several authors \citep{Jiang2013, Amari2014, Yamasaki2022, Zhong2023} showed that in data-constrained simulations of active regions, eruptions are triggered by torus instability and, magnetic reconnection allows the full development of the instability  by removing part of the line-tied connections.
Another approach is to start from an initial stable magnetic configuration and trigger the eruption through photospheric motions, such as shearing and convergence \citep{Inoue2018, Torok2018}, or flux emergence \citep{Fan2011, Fan2022, Muhamad2017, Inoue2025}. 

On the other hand we also have fully data-driven MHD simulations, where the bottom boundary conditions are specified by the observables at the photosphere in a time-dependent way, covering from several hours to a couple of days of the pre-eruption phase \citep{Jiang2016, Wang2023, Chen2023, Guo2023}. Since it is computationally expensive to simulate the long-term pre-eruption phase in full MHD, another option is to use the data-driven magnetofrictional technique \citep{Cheung&DeRosa2012, Gibb2014, Pomoell2019, Price2019, Price2020, Kilpua2021} to simulate the process ahead of the eruption, and apply the full MHD modelling only to the eruption itself \citep{Afanasyev2023, Daei2023, Guo2024}. However, choosing an appropriate time step for switching from magnetofriction to MHD is non-trivial as the properties of the erupting system significantly depend on the chosen simulation frame for this transfer \citep{Wagner2024c}.

In all the data-driven or data-constrained simulations mentioned above, the coronal fields are numerically extrapolated from the photospheric magnetic distribution based on a force-free assumption, i.e., no Lorentz force is exerted \citep{Wheatland&Gilchrist2013, Wiegelmann&Sakurai2021}.  
This is based on the hypothesis that the coronal magnetic fields evolve continuously driven by motions at the photosphere in a quasi-static way.
Although simulations based on NLFFF extrapolations are capable of reproducing realistic dynamics, the magnetic field in the photosphere is generally not perfectly force-free \citep[e.g.][]{Schuck2022}. Also, the force-free approximation is based on the assumption of negligible plasma-$\beta$, but this is only true in the transition region and low corona \citep{Gary2001}.
Since the extrapolations for the aforementioned models take their bottom boundary in the photospheric magnetic field below the chromosphere, where $\beta \geq 1 $, a non-force free extrapolation approach should be generally more realistic. 

Thus, let us consider a non-force-free field (NFFF hereafter) extrapolation, prescribing the magnetic field at the photosphere with a vector magnetogram close to the onset of the flare. As the name suggests, this technique allows to have a non-zero Lorentz force, which is a crucial point because during solar eruptions the coronal magnetic field is far from equilibrium. The NFFF extrapolation relies on self-organized double-curl Beltrami fields \citep{Bhattacharyya2007} satisfying the minimum dissipation rate principle \citep{Bhattacharyya&Janaki2004, Bhattacharyya2007, Hu&Dasgupta2008}. Comparing with the widely used NLFFF \citet{Agarwal2022} have shown that the amount of free magnetic energy found in an active region simulation initialised with NFFF extrapolation is larger compared to the simulation using NLFFF extrapolation (2.5 times larger). In addition, NFFF shows a slightly better correlation with the measured photospheric field, since for the force-free extrapolation a ``preprocessing'' technique is often performed on the data to remove the Lorentz force and obtain a suitable boundary condition \citep{Wiegelmann2006, Jiang&Feng2014}. Furthermore, for the NFFF the Lorentz force at the photosphere is non-zero but generally decays sharply with height, making it approximately force-free in the corona \citep{Agarwal2022}, as it is expected for the Sun. 
\citet{Prasad2017} showed that the inherent non-zero Lorentz force from NFFF extrapolations can trigger an eruption and successfully reproduce the coronal dynamics that trigger solar flares \citep{Prasad2018,Kumar2022,Kumar2025}, coronal jets \citep{Nayak2019, Joshi2024}, and coronal dimmings \citep{Prasad2020}.

\begin{figure*}[!ht]
  \centering 
  \includegraphics[width=\textwidth]{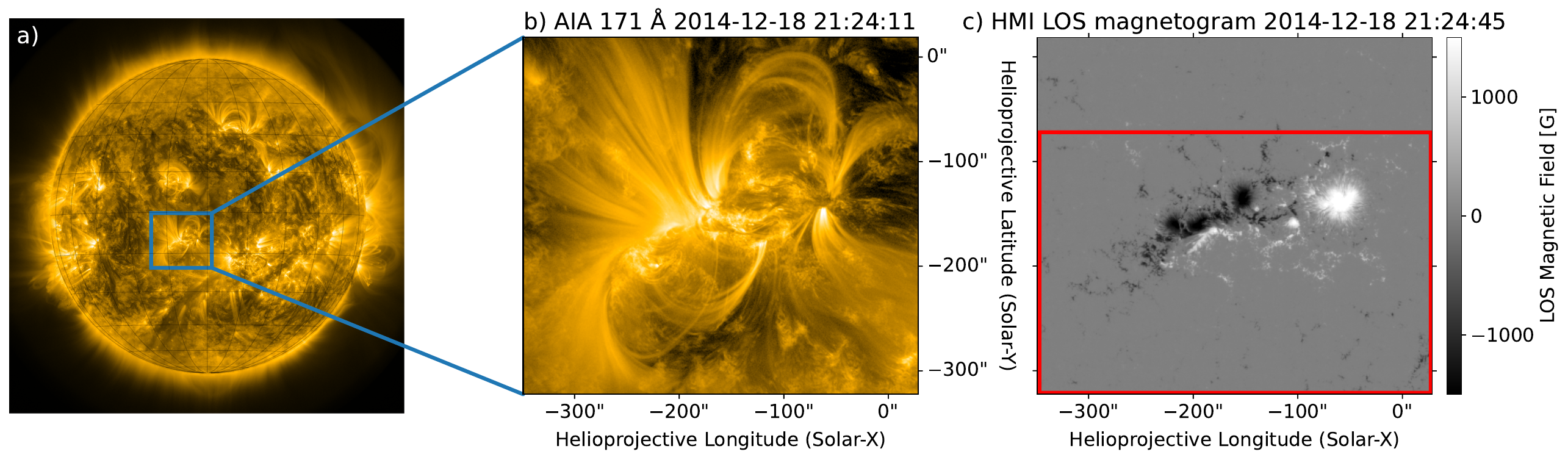}
  \caption{Active region NOAA 12241. a) Full-disk image in 171~\AA{} observed by SDO/AIA on December 18, 2014 at 21:24 UT. The blue rectangle indicates the active region under study. b) Zoom-in of the active region in 171~\AA{}. c) Line-of-sight (LOS) magnetogram taken by SDO/HMI. The red rectangle indicates the horizontal domain for the numerical simulation.}
  \label{fig:obs}
\end{figure*}

Considering the NFFF extrapolation together with a gravitationally stratified atmosphere, ranging from the photosphere to the corona, we have developed a fully resistive and compressible MHD numerical framework to study solar eruptions in active regions, including optically thin radiative cooling and thermal conduction. As a case study, we consider the eruption from AR 12241 on December 18, 2014. This event was previously modelled by \citet{Prasad2023} using an incompressible regime and an isothermal corona. 
They found that a flux rope formed due to the initial converging Lorentz force, which drove tether-cutting reconnection below the rope and led to its slow expansion. In the present study, we also find that a flux rope self-consistently develops and rises, carrying chromospheric material. In addition, in our model the flux rope does not stop its expansion but rather is expelled out of the simulation box in a second phase after a first slow rise. With the aid of an algorithm that identifies and tracks this structure \citep[see][and references therein]{Wagner2024a, Wagner2024b}, we study the dynamics of the magnetic flux rope and determine its kinematic properties, finding a great similarity with observations.

The article is organised as follows: in Sect.~\ref{s:digest} we introduce the observations of the active region, followed by a summary of the MHD simulation results. Then in Sect.~\ref{s:FRFormation} we describe the initial stage of the simulation, including the compression and formation of the flux rope, and in Sect.~\ref{s:Rising} we focus on the evolution including a force analysis of the rising phase. Finally in Sect.~\ref{s:disc_and_concl} we outline a discussion comparing with other models and the main outcomes of our study. 
All the numerical framework, together with the initial conditions and details on the flux rope identification and tracking algorithm are presented in the appendix (Sect.~\ref{s:setup} and \ref{s:guitar}). Complementary material to Section~\ref{s:Rising} is added in Appendix~\ref{s:risingsuppl} including the evolution of all the forces involved in the rising phase.

\section{Modelling overview of active region 12241}
\label{s:digest}

In this section we introduce the eruptive event under study together with the active region where the eruption takes place. We describe briefly the extrapolation method applied to determine the magnetic field in the solar atmosphere that we use later as initial condition of our compressible MHD numerical framework.
We present also the dynamics achieved in the simulation, summarizing and highlighting the main outcomes of it, in order to provide a global picture before proceeding to a more detailed description in Sect.~\ref{s:FRFormation} and \ref{s:Rising}.

\subsection{Eruptive event}
\label{ss:observations}

We consider observations of NOAA AR 12241 on December 18, 2014 at 21:24~UT, shown in Fig.~\ref{fig:obs}. Panel a) in the Fig. displays the full-disk image of the Sun in 171~\AA{} taken by the Atmospheric Imaging Assembly \citep[AIA,][]{aia} on board the Solar Dynamic Observatory \citep[SDO,][]{sdo}. The blue rectangle highlights the active region under study. A zoom-in of this region is shown in panel b), with the accompanying photospheric line-of-sight (LOS) magnetic field in panel c), taken by the Helioseismic Magnetic Imager \citep[HMI,][]{hmi} also on board SDO. The red rectangle on panel c) indicates the numerical domain used in this work. Since the eruption is heading southwards, we chose an extended domain in this direction only, to maximize numerical efficiency. This event was previously observationally analysed in detail by \citet{Joshi2017}. They described it as a three-ribbon flare with a two-stage eruption. The first stage consists of the formation and rise of a flux rope above the polarity inversion line due to a tether-cutting reconnection process, displaying the characteristic two-ribbon flare phenomenology according to the standard model. The second part comprises reconnection of the overlapping arcade containing the magnetic flux rope at a higher three-dimensional null-point, resulting in a quasi-circular ribbon shape. The ﬂare produced by this active region was class M6.9 on the GOES X-ray scale. It started at $\sim$21:41~UT, peaked at $\sim$21:58~UT, and the decay phase continued until $\sim$02:00~UT on December 19, 2014. 

\begin{figure*}[t!]
  \centering
  \includegraphics[width=\textwidth]{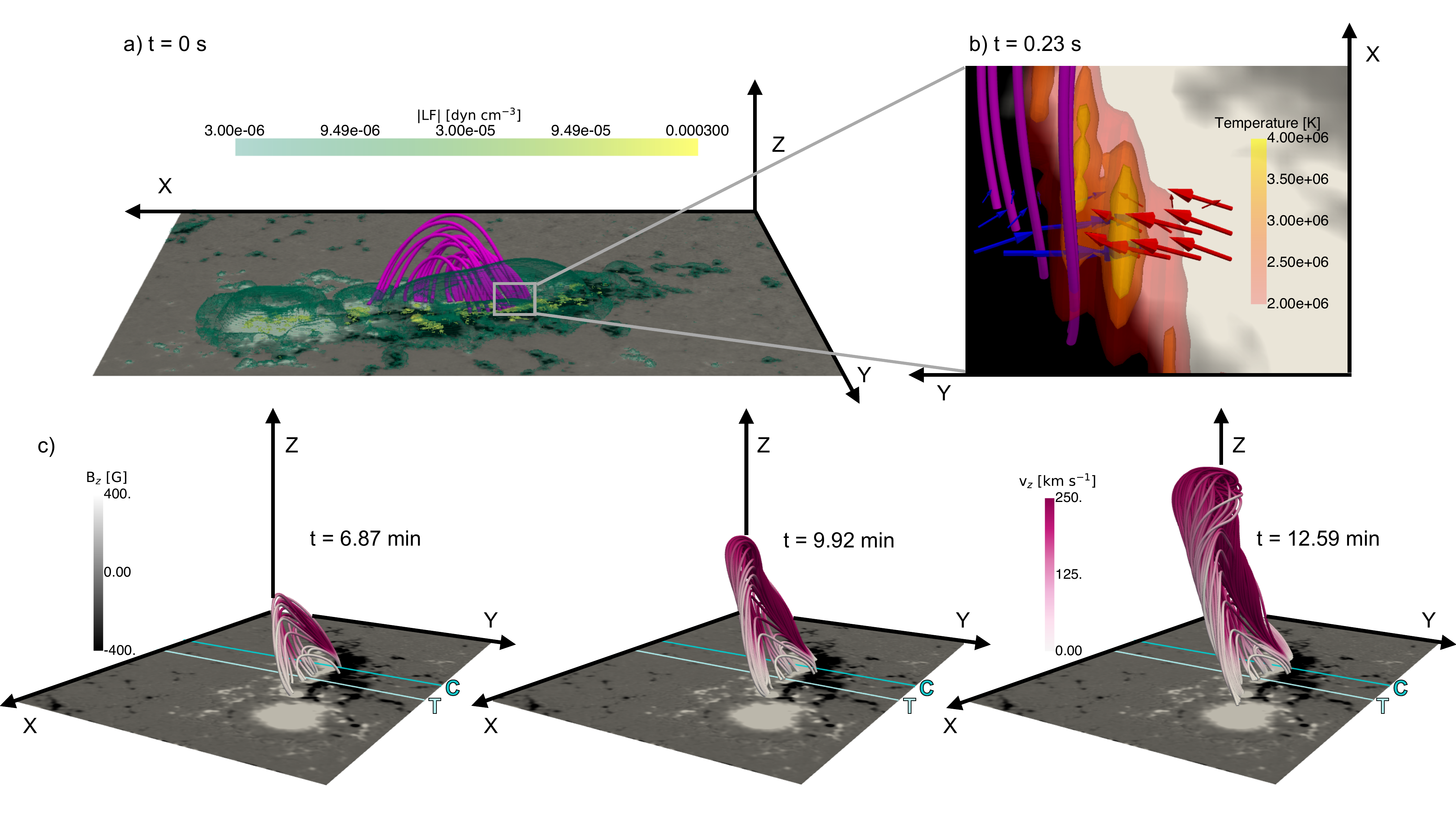}
  \caption{
  Initialization and evolution of the eruption at glance. In all the panels, a magnetogram, with white to black levels, is present in the background to make the link with observations in Fig.~\ref{fig:obs}. The colour map is saturated, between -400 and 400~G for visualization purposes.
   a)  Initial magnetic configuration and Lorentz force at $t=0$~s. The pink lines represent the magnetic arcade which later on is transformed in the erupting flux rope. The green to yellow colour map indicate a volumetric representation of the magnitude of the Lorentz force, being yellow (green) more (less) intense. 
   b) A zoom-in at $t=0.23$~s within the region defined by the grey rectangle of panel a). The horizontal components of the Lorentz force are shown with arrows in blue (negative) and red (positive) drawn at height $z=2.2$~Mm.
   Temperature iso-surfaces of $2 \times 10^6$~K (red) to $4 \times 10^6$ (yellow) are also displayed. 
   c) Rising phase is depicted for $t=6.87$~min, $t=9.92$~min and $12.59$~min. The coloured lines represent the magnetic field lines corresponding to the flux rope. The colour of the field lines represents the vertical speed $v_z$ in pink.
   The two lines on the $xy$-plane indicate the cuts where the tracking (T line in light turquoise) of the flux rope and the compression (C line in dark turquoise) analysis are performed, respectively.}
  \label{fig:3D}
\end{figure*} 

\subsection{Modelling the dynamics of AR 12241}
\label{ss:extrapolation}

To model the magnetic field of the active region, we use a NFFF extrapolation as initial condition. Briefly, this method is based on the principle of minimum energy dissipation rate, in which the magnetic field has to satisfy the double-curl Beltrami equation \citep[see ][]{Bhattacharyya2007}. Then the magnetic field $\mathbf{B}$ can be written as $\mathbf{B}=\mathbf{B}_1+\mathbf{B}_2+\mathbf{B}_3$,
where $\nabla \times \bf{B}_{1,2}=\alpha_{1,2} \bf{B}_{1,2}$ correspond to two linear force-free fields and $\bf{B}_3$ with $\alpha_3=0$ corresponds to a potential field \citep{Hu&Dasgupta2008}. The parameters $\alpha_{1,2}$ are normalized by the horizontal grid size and they are chosen to minimise the value of the difference between the computed and measured transverse magnetic field vectors at the bottom boundary. For this AR the values are $\alpha_1=0.0105$ and $\alpha_2=-0.0105$. The units of $\alpha$ are pixel inverse with a pixel size of 1 arcsec. 
For more details about the model, readers are referred to \citet{Hu2010} and references therein. The code to calculate a NFFF extrapolation for active or quiet regions is available in \url{https://github.com/AvijeetPrasad/MDR-NFFF/}. The linear combination of two linear force-free fields and one potential field leads the total field to be non force-free (hence the name, non force-free field extrapolation). As we mentioned in Sect.~\ref{sec:intro}, force free extrapolation generally requires an external driver  (e.g. photospheric flows, flux emergence) to trigger the eruption. In an NFFF extrapolation, the initial state is imbalanced and this inherent force can drive an eruption \citep[see][]{Prasad2017,Prasad2023}. Previous experiments considering the same active region using earlier magnetograms, not shown in this paper for the sake of conciseness, indicated that such an initial imbalance state can as well lead to a quiet relaxation of the system, without necessarily leading to an eruptive event.

In the case presented in this work, the NFFF initial condition induces the self-consistent development of a flux rope, that subsequently rises and leaves the computational domain. This result is obtained by inserting the NFFF into a plane-parallel stratified atmosphere structured like the low atmosphere of the Sun. It includes a dense and cold lower layer akin to a chromosphere, a narrow transition region and an extended and hot corona (for more details on the initial conditions see Fig.~\ref{fig:init_atm} and Fig.~\ref{fig:B_init} from Appendix~\ref{ss:ic}). The compressible MHD equations are then solved with the PLUTO code \citep{Mignone2007} to study the evolution following this initial imbalance, including thermal conduction as well as radiative cooling. We defer the detailed description of the MHD modelling to Appendix \ref{s:setup}. More precisely, we describe the MHD equations used in Appendix~\ref{ss:eq}, the numerical grid in Appendix~\ref{ss:grid}, the initial conditions in Appendix~\ref{ss:ic}, and the boundary conditions in Appendix~\ref{ss:bc}. We next focus on describing the two stages of dynamical evolution obtained with our model. 

\subsubsection{Flux rope formation and mass loading} 
A summary of the evolution of the simulation is shown in Fig.~\ref{fig:3D}. The top panels display the initial stage and the bottom panels present the evolution of the rising structure. Initially the plane parallel atmosphere is in hydrostatic equilibrium but the non-force-free magnetic field is imbalanced. Panel a) shows at $t=0$~s the magnetic field lines of an arcade (magenta lines) that are going to evolve into the eruptive flux rope and the non-zero Lorentz force as green/yellow iso-surfaces).  
Strong Lorentz force concentrations are found especially in the lower atmosphere (highlighted by the yellow patches), near the arcade foot points. Thus, the out of equilibrium magnetic field destabilizes the atmosphere, inducing motions and subsequent compression and heating of the plasma in the upper part of the transition region. This is shown in a small region highlighted by a grey square in panel b). The horizontal component of the Lorentz force (blue and red arrows) rapidly ($t=0.23$~s) induces a flow that leads to compression in the upper part of the transition region (around $z=2.5$~Mm), close to the highlighted magnetic arcade. This compression leads to a local increase of the temperature by a factor of 8 (yellow volumes at $4 \times 10^6$~K in panel b). Thermal conduction then takes over, heating up the top of the transition region from above and eventually leading to evaporation. This evaporation fills up the magenta arcade with denser material, setting up the mass loading.

After evaporation is initiated, reconnection between opposite legs of the arcade are starting to take place (see Fig.~\ref{fig:sq} from Appendix~\ref{s:reconnection} for a quantitative description). This reconnection increases the twist of the arcade, leading to the formation of the flux-rope.

\subsubsection{Rise and escape of the flux rope} 
\begin{figure}[!ht]
    \centering
    \includegraphics[width=\linewidth]{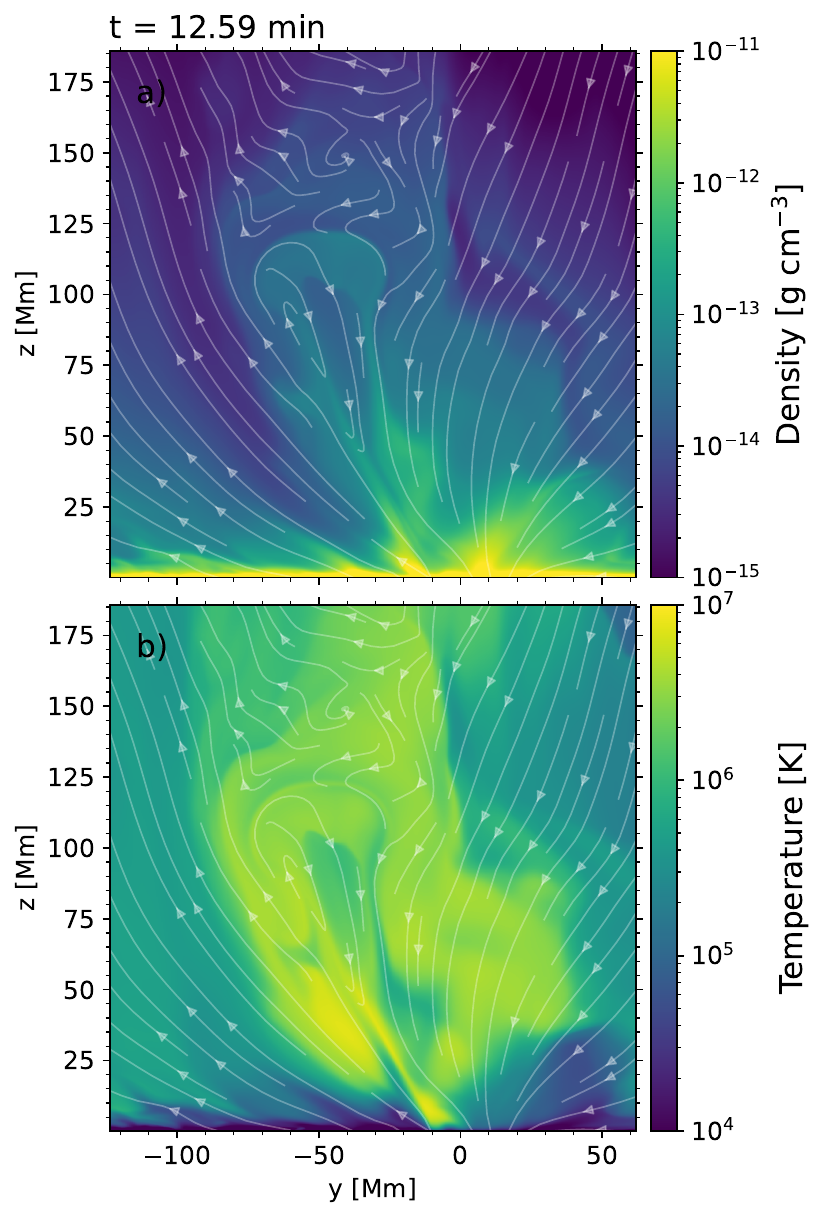}
    \caption{Thermodynamical variables in the rising phase. a) Density and b) temperature at $x=5$~Mm (corresponding to line T in Fig.~\ref{fig:3D}~c)) for $t=12.59$~min. Yellow denotes higher density and temperature and blue lower values. White solid lines represent the streamlines of the projected magnetic field vector on the plane. A movie showing the temperature in this cutting plane together with 3D visualization of the flux rope field lines is available in the online version.}
    \label{fig:rho_temp}
\end{figure}

Later on in the simulation, see Fig.~\ref{fig:3D}~c), the flux rope expands and rises. 
As the flux rope rises, it carries up dense and hot material. This is shown in Fig.~\ref{fig:rho_temp}, where panel~a) exhibits the density and panel b) the temperature in a $yz$-plane at $x=5$~Mm for $t=12.59$~min. This plane  corresponds to the T line (light turquoise) from Fig.~\ref{fig:3D}~c) and intersects the flux rope transversely near its the apex. Note the denser and hotter plasma (in yellow) compared with the surrounding corona, delimited by ellipse-shaped field lines of the projected magnetic field onto the plane (white streamlines representing the magnetic field lines on that plane). 

\begin{figure}[!ht]
    \centering
    \includegraphics[width=\linewidth]{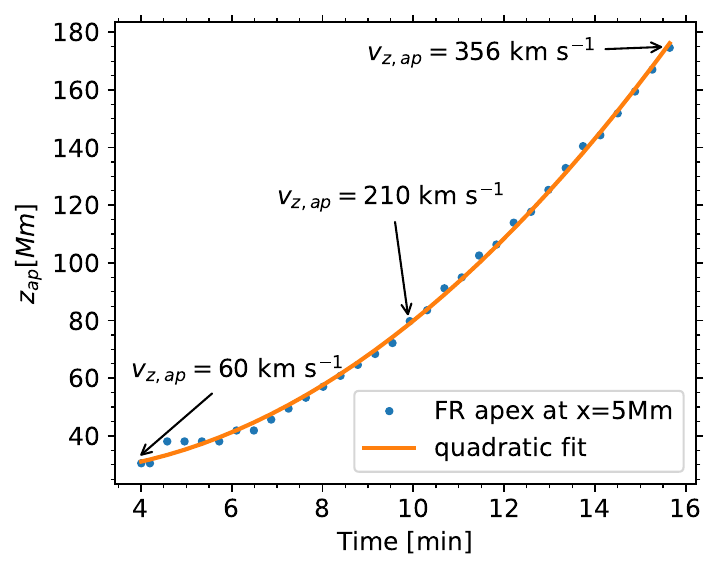}
    \caption{Kinematic properties of the flux rope apex. Vertical coordinate of the flux rope apex at $x=5$~Mm (blue dots) and quadratic fit (solid orange line). Values of the vertical velocity are added to different times. Before $t=4$~min the flux rope is not sufficiently well formed to infer the apex height.}
    \label{fig:FRap}
\end{figure}

Around $t=6$~min, the interplay between the vertical gas pressure gradient and the Lorentz force gives a positive net contribution, and, as a consequence, the fast rising motion is more evident (see Fig.~\ref{fig:FRap})
This figure shows in blue dots the $z$-coordinate of the apex of the flux rope identified in the same cutting plane shown in Fig.~\ref{fig:rho_temp} together with a quadratic fit (orange solid line). The apex is tracked until it leaves the domain at $t=16$~min with a speed around 350~km s$^{-1}$. The identification and the tracking of the flux rope will be explained in detail in Sect.~\ref{ss:FRid}. The legends with the arrows display the vertical speed of the flux rope apex $v_{z,ap}$ at $x=5$~Mm obtained from the derivative of the fitting curve at a few selected times. As a consequence of the quadratic height profile, we obtain a vertical constant acceleration of $424$~m s$^{-2}$, which is 1.5 larger than the gravity at the solar surface.

\begin{figure}[!ht]
    \centering
    \includegraphics[width=\linewidth]{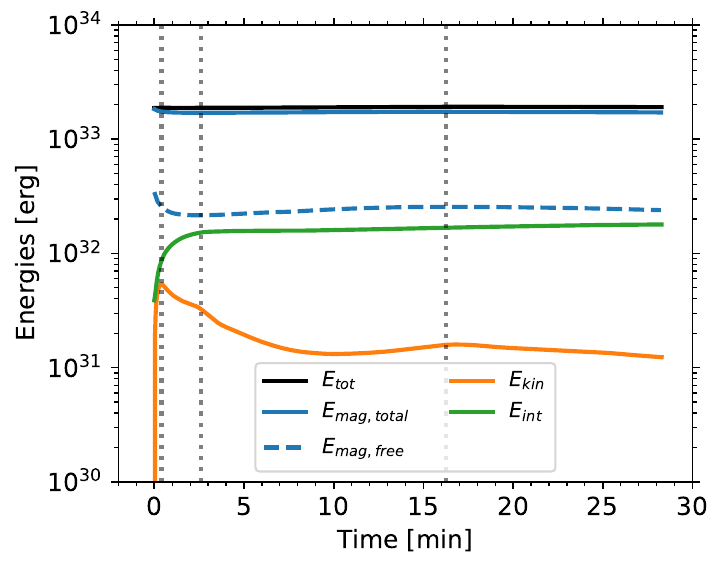}
    \caption{Evolution of energies averaged in the entire computational domain. Total energy is in solid black line, total magnetic energy in solid blue line, free magnetic energy in dashed blue, kinetic and internal in orange and green solid lines, respectively. Vertical dotted lines indicate different stages. }
    \label{fig:energy}
\end{figure}

\begin{figure*}[!ht]
    \centering
    \includegraphics[width=\textwidth]{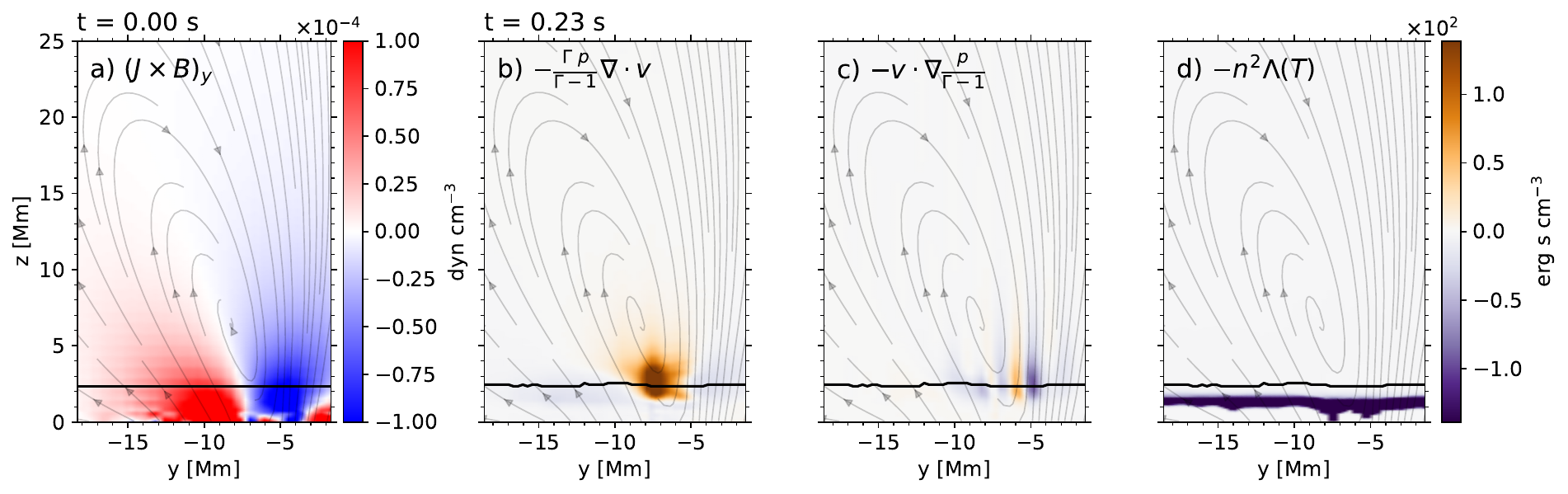}
    \caption{Lorentz force and energy source terms for the plane $x = -15$~Mm (see its photospheric trace C in Fig.~\ref{fig:3D}). a) Lorentz force in the y-axis at $t = 0$~s (positive values are in red and negative in blue). b) Compression, c) advection and d) radiative cooling at $t = 0.23$~s (positive values are indicated in dark orange and negative in purple). The upper limit of the transition region is denoted with a black solid line.}
    \label{fig:e_src}
\end{figure*}

\subsubsection{Energetics} 
To better understand the dynamics observed in our simulation, we complement the previous simplified analysis with the overall evolution of the energy content in our simulation box, which is shown in Fig.~\ref{fig:energy}. The total energy (black solid line) remains roughly constant, meaning that the losses (or gains) through the boundaries are small compared to the total energy content in the simulation box. The total magnetic energy (blue solid line) is similar to the total energy, showing that most of the energy in the domain is due to the magnetic field. From the calculation of the potential field (following the method proposed by \citealt{Alissandrakis1981}), we estimate the magnetic energy associated with it and the free magnetic energy (blue dashed line) as $E_{\rm mag,free}=E_{\rm mag,tot}-E_{\rm mag,pot}$. Note that $E_{\rm mag,pot}$ is almost constant in time as the magnetogram is not evolving (it diminishes slowly over the course of the simulation due to diffusion at the bottom boundary). This decrease in the potential energy is very small and does not affect the free magnetic energy evolution.   
The free magnetic energy originated from the NFFF extrapolation initially has a value of $\sim 2\times 10^{32}$~erg and then diminishes because of ohmic dissipation, reaching a minimum at $t=2.6$~min (second dotted vertical line). This decrease corresponds to an increase in the internal energy (green solid line) and a sudden increase of the kinetic energy (orange solid line) up to $t=0.40$~min (first dotted line).
After $t=2.6$~min the free magnetic energy slowly increases up to $t=16.3$~min (third vertical dotted line) as the plasma motions lead to the increase of the magnetic strength of the arcades.  
This rise is compensated with a decrease of the kinetic energy, mostly due to the work of the magnetic tension and, in a smaller proportion, due to other high-speed mass leaving the domain on the sides or through the top boundary. 
However, after $t=10$~min, there is a slight increase in the kinetic energy, up to $t=16.3$~min. This is likely caused by the accelerated flux rope that is ascending until it leaves the domain at $t=16.3$~min.
After this time, both free magnetic and kinetic energies smoothly drop, due to magnetic relaxation and the departure of the magnetic flux rope out of the simulation box.

In the next sections we explain more in detail the different phenomena involved in the formation of the flux rope (Sect.~\ref{s:FRFormation}) and its rising phase (Sect.~\ref{s:Rising}). 
\section{Flux rope formation and plasma evaporation}
\label{s:FRFormation}

In this section we describe the initial stage of the simulation, which is crucial in the triggering of the eruption. Since the dynamics is very fast at the beginning, we run the simulation using high temporal cadence outputs, every 0.11~s, to study the evolution of all the variables and the physical processes involved. 

\begin{figure*}[!ht]
    \centering
    \includegraphics[width=\textwidth]{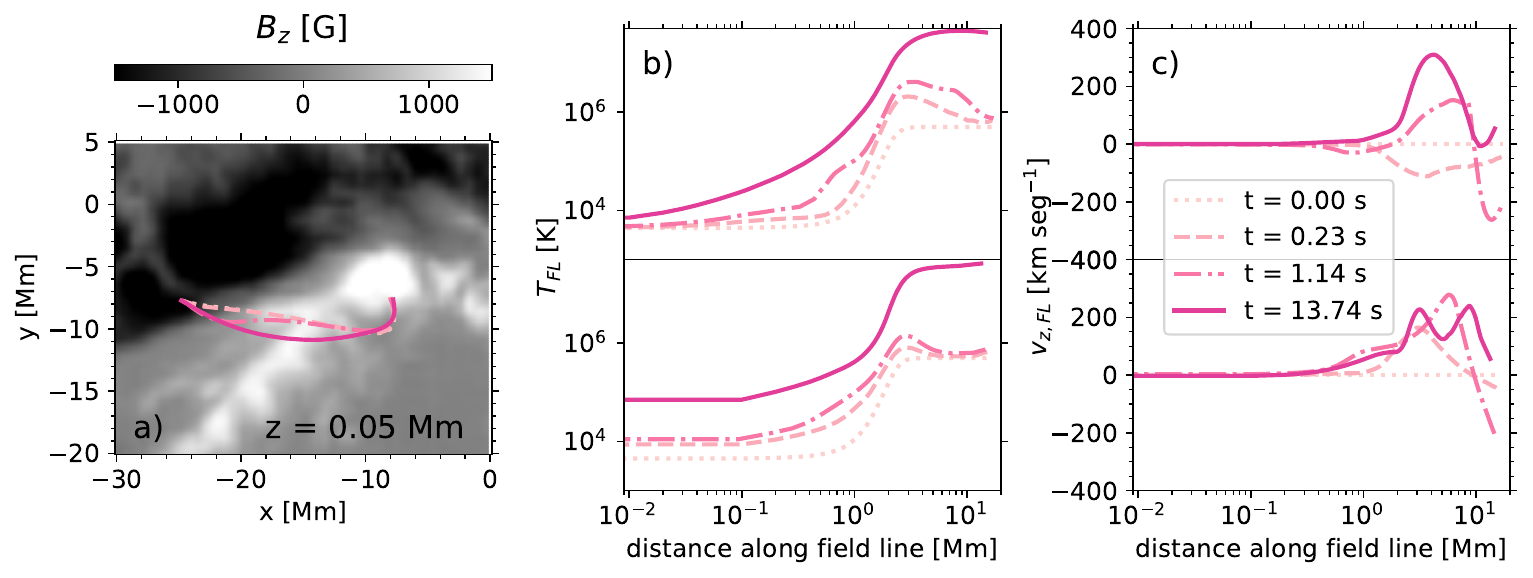}
    \caption{Evolution of thermodynamical variables along a magnetic field line. a) Plane-of-sky coordinates of a magnetic field line, represented with dotted ($t=0$~s), dashed ($t=0.23$~s), dashed-dotted ($t=1.14$~s), and solid lines ($t=13.74$~s), from light to dark pink. On the background $B_z$ at $z=0.05$~Mm and $t=0$~s is shown in grey scale, with negative values in black and positive values in white. b) Temperature and c) vertical velocity along the field line from the left footpoint to the apex (top panels) and from the right footpoint to the apex (bottom panels). The variables are plotted for the same instants of time as in panel a).}
    \label{fig:tempFL}
\end{figure*}

\subsection{Role of the initial Lorentz force}
\label{ss:Compression}

Given the non force-freeness of the extrapolation, the initial Lorentz force is out of equilibrium ($\bf{J} \times \bf{B} \neq 0$). In particular, its horizontal components along the $x$ and $y$ axis are pointing in opposite converging directions near the onset of the eruption, around $x=-15$~Mm, $y=-7$~Mm, in the vicinity of the polarity inversion line. This imbalance of the Lorentz force triggers a converging flow leading to compression of the plasma in the upper part of the transition region in a fraction of seconds. This is shown in Fig.~\ref{fig:e_src}, where panel a) displays one of the horizontal components of the Lorentz force $(\mathbf{J} \times \mathbf{B})_y$ at a plane corresponding to $x=-15.14$~Mm (turquoise line C in Fig.~\ref{fig:3D} panel c)) in a small region of the computational domain. The direction of the force is shown by the negative (blue) and positive (red) values. 

Panel b) of Fig.~\ref{fig:e_src} shows the compression term at $t=0.23$~s for the same plane defined by the line C. This term is obtained from the combination of the flux of the internal energy and the work of the pressure gradient from the internal energy equation and is proportional to $\nabla \cdot \mathbf{v}$ (see Eq.~\eqref{eq:int_energy}). 
The purple corresponds to negative contribution and orange to positive values. 
Panels c) and d) exhibit the contribution of the internal energy advection term and the radiative cooling, respectively, for the same time and plane. $\Gamma=5/3$ is the ratio of specific heats and $p/(\Gamma -1)$ corresponds to the internal energy, while $n$ is the density number and $\Lambda$ is the radiative loss function defined in Sect.~\ref{ss:eq} by Eq.~\eqref{eq:rad}. The grey lines on the background of all panels correspond to the integral lines of the field components contained in the plane $x=-15.14$~Mm. The horizontal solid black lines denote the upper limit of the transition region, i.e. the beginning of the corona. To calculate it we search for the minimum of the second derivative of the temperature with respect to height for each value of $y$.

From panel b) it is noticeable that the localized compression between $z=1.5$~Mm and $z=4.5$~Mm and around $y=-7.5$~Mm, is at the same location where $(\mathbf{J} \times \mathbf{B})_y$ converges. Panel c) illustrates that the advection term is contributing with both positive and negative effects, although in smaller amounts. Moreover, the advection term is mostly important on the right side of the flux rope.  Finally, panel d) indicates that radiative cooling is removing mostly energy from the system from below the transition region as expected. 

As a consequence of the initial compression due to converging flows triggered by the initial distribution of the Lorentz force, the density and the temperature increase locally with time. To illustrate this effect in more detail, we present in Fig.~\ref{fig:tempFL} b) the temperature profile as a function of the distance for a single field line that is part of the flux rope, for different times. The times are represented by different line styles. 
The chosen field line has undergone marginal slippage during the times analysed at this early stage. To verify this, we have computed the squashing factor (not shown here for brevity) on the bottom plane where the field lines are rooted. For the positions of the left foot point we obtained values of $|\log(Q)|\sim1$ and for the right one, $|\log(Q)|\sim2$. These values are small compared to $|\log(Q_{max})|\sim6$ for this plane,  suggesting that there is no important field line slippage along the field line at this time interval for these locations. Indeed, there is negligible displacements of the field line foot points. Next, to highlight the process in the transition region, the field line is divided in two parts, one from the left foot point to the apex (top panel) and the other from the right foot point also to the apex (bottom panel). The increase of temperature over time to both sides of the apex of the field line is very noticeable, from $5 \times 10^5$~K to $5 \times 10^7$~K, especially at the lower corona, where the compression occurs. The $x$ and $y$ coordinates of this field line at the different times are shown in panel a), together with $B_z$ at $z=0.05$~Mm in the background.
The method used to identify and track the flux rope will be explained in Sect.~\ref{ss:FRid}.   

\subsection{Thermal conduction and evaporation}
\label{ss:Evaporation}

Immediately after compression begins, there is an increase in temperature, as previously described,and a downflow on both sides of the region that is being heated. 
Because of temperature rise at the apex, the thermal conduction flux grows in time along the magnetic field lines, heating up the plasma at the bottom of the transition region. As a consequence some hot material starts moving upwards of the transition region, i. e., evaporating. In Fig.~\ref{fig:tempFL}~c) we present the vertical velocity $v_z$ along the field line for the same time steps described in panels a) and b). The field line is also divided in two parts to see in more detail the evaporation above the transition region. At $t=0.23$~s (dashed line) there is a flux moving downwards (top panel) near the location where the plasma is heated (see the corresponding peak in temperature in panel b), around 2 and 3 Mm. 

Later on, after the plasma has been heated, around $t=1.14$~s (dashed-dotted line), there is a flux going up from the transition region to the corona, and, from $t=13.74$~s onward this upflow remains and becomes larger on one side of the field line (left footpoint). This upflow is responsible for loading up the whole flux rope with mass during these initial stages of the simulation.

\section{Flux rope rise and deflection}
\label{s:Rising}

\begin{figure*}[!ht]
    \includegraphics[width=\textwidth]{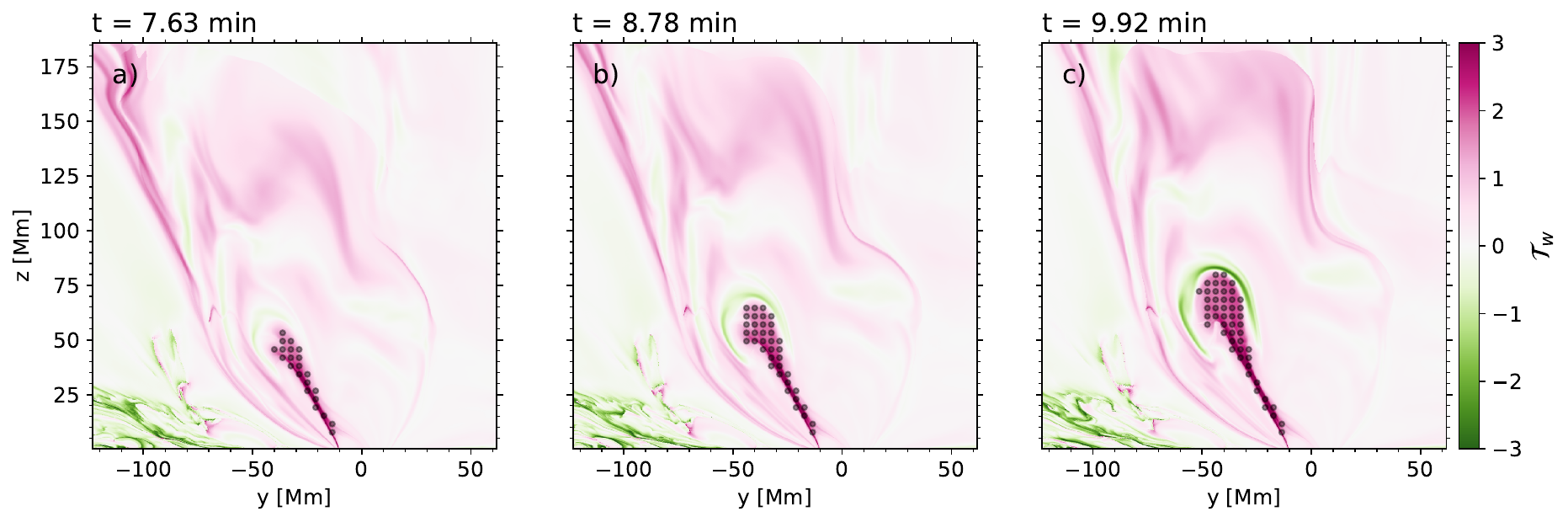}
    \caption{Twist $\mathcal{T}_w$ at $x=5$~Mm (cut T in Fig.~\ref{fig:3D}) for a) $t=7.63$~min, b) $t=8.78$~min and c) $t=9.92$~min. Pink indicates positive values of twist related with a current density parallel to the magnetic field for the corresponding field line and green denotes anti-parallel orientation. The grey circles represent the points belonging to the flux rope extracted with GUITAR.}
    \label{fig:twist}
\end{figure*} 
 
As evaporation and reconnection persist in the lower part of the flux rope, the magnetic field lines belonging to it are loaded with mass while they continue expanding and twisting. At about $t=4$~min the flux rope is well-defined and starts to rise slowly, driven by a net upward force, combination of the Lorentz force and the gas pressure gradient. At the same time the structure is moving upwards, it is deflected to the southeast, in the same direction as in observations, towards a region of low magnetic pressure. In this section we will describe the dynamical analysis of the forces acting on the flux rope in the rising phase.

\begin{figure*}[!ht]
    \includegraphics[trim={0cm 0cm 0cm 8cm},clip, width=\textwidth]{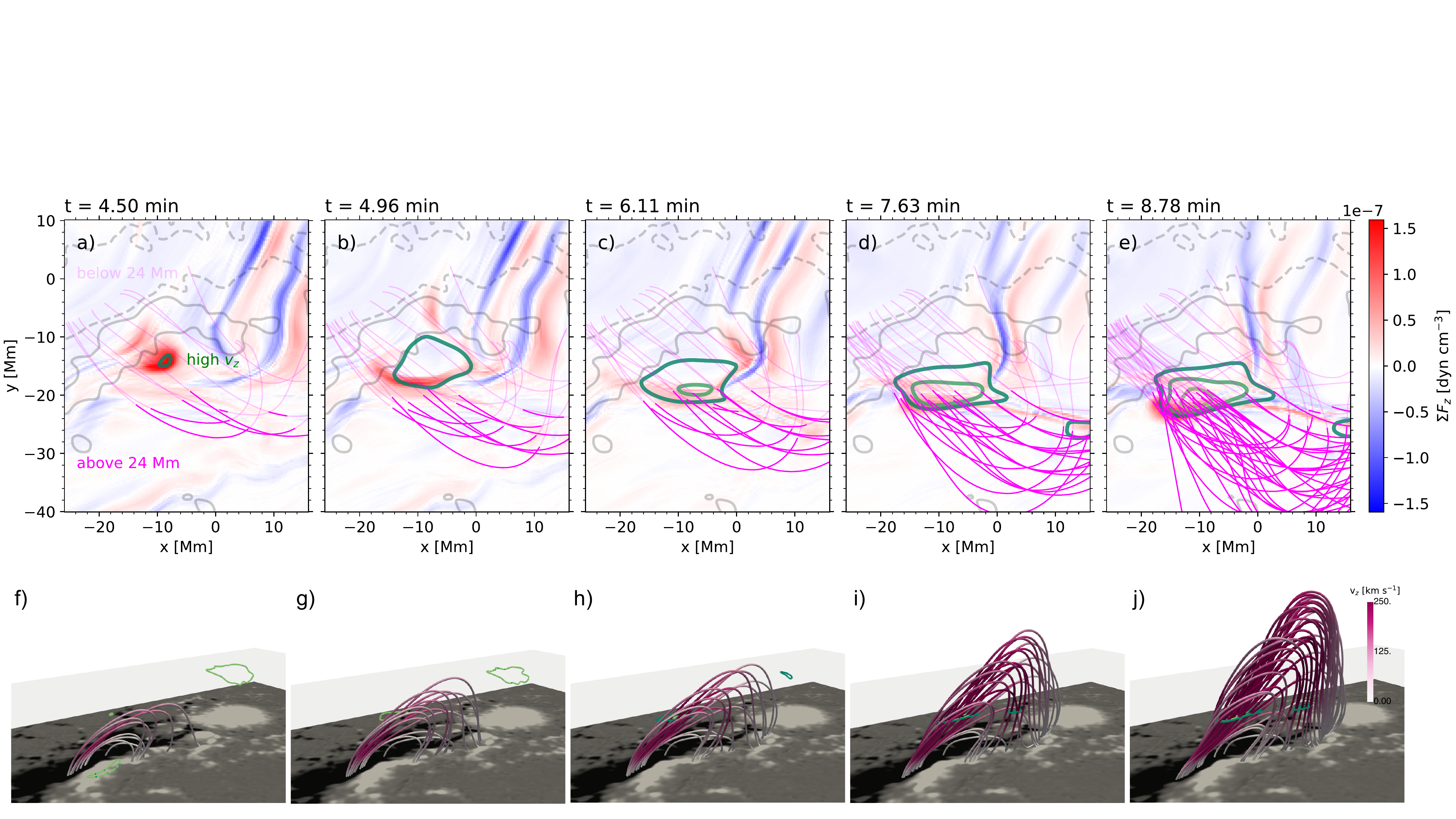}
    \caption{Top panels: Evolution of the net vertical force  (positive values are in red and negative values in blue) at $z=24$~Mm for a) t=4.5 min, b) t=4.96 min, c) t=6.11 min, d) t=7.63 min and e) t=8.78 min. The magenta lines correspond to  field lines computed within the flux rope. The grey contours indicate positive (solid) and negative (dashed) values of $B_z$ at $z=0.05$~Mm and the green contours exhibit different values of $v_z$ between 218 (dark green) and 436 km s$^{-1}$ (light green). Bottom panels: Flux rope magnetic field lines in 3D coloured by $v_z$ for the same times as top panels. The grey semi-transparent plane corresponds to $z=24$~Mm and contains the same $v_z$ contours as the top panels. }
    \label{fig:netFz_zcut}
\end{figure*}

\subsection{Flux rope identification}
\label{ss:FRid}

For identifying and tracking the flux rope in time we use GUITAR \citep[Graphical User Interface for Tracking and Identifying flux Ropes,][\url{https://github.com/wandi0909/GUITAR}]{Wagner2024a, Wagner2024b}. The method uses combinations of mathematical morphology algorithms, such as erosion and dilatation, to extract the magnetic flux rope throughout a time series. More details about the method are given in App.~\ref{s:guitar}. 
A flux rope is a group of helical field lines collectively winding around a common axis \citep{Liu2020}.
The winding number $\mathcal{T}_g$ \citep[see Eq. 12 in ][]{BergerandPrior2006} quantifies for each field line the number of turns made around the flux rope axis. For 3D magnetic configurations, one of the difficulties for calculating $\mathcal{T}_g$ is to find the flux rope axis. An easier quantity to compute is the twist number defined by 
    $$\mathcal{T}_w=\int_L \frac{\nabla \times \mathbf{B} \cdot \mathbf{B}}{4 \pi B^2} dl\, ,$$
where $\nabla \times \mathbf{B}$ is the current density, $dl$ is the elementary length along the field line and $L$ the total length of the studied field line. $\mathcal{T}_w$ supplies the number of turns that two infinitesimally adjacent field lines make around each other \citep{BergerandPrior2006}. $\mathcal{T}_w$ is an approximation of $\mathcal{T}_g$. 
More precisely, $\mathcal{T}_w$ approaches $\mathcal{T}_g$ in the vicinity of the axis of a nearly cylindrically symmetric flux tube \citep{Liu2016}. 
For determining the location of the flux rope's axis assumptions are necessary and often involve the computation of $\mathcal{T}_w$, where the axis is assumed to be located at a local extrema of $\mathcal{T}_w$ \citep{Liu2016, Price2024}. In contrast, $\mathcal{T}_w$ can be computed straightforwardly for any field line and for the case of thin flux ropes errors are rather small \citep[for more details see][]{Liu2016,Liu2020}.

We apply GUITAR to $\mathcal{T}_w$ calculated along the field lines that go through a 2D plane located at $x=5$~Mm (cut T in Fig.~\ref{fig:3D}), shown in Fig.~\ref{fig:twist}. This figure displays the evolution of $\mathcal{T}_w$ for a) $t=7.63$~min, b) $t=8.78$~min and c) $t=9.92$~min in green (negative) to pink (positive) colour map. The grey circles correspond to the location of the flux rope extracted with GUITAR. Note that the flux rope is associated with a highly coherent twisted structure that remains connected to the photosphere during the entire rising phase until it leaves the domain. As the flux rope moves upward, a current layer of opposite twist number ($\mathcal{T}_w<0$) develops at its boundary.

The chosen plane cuts transversely the flux rope through the apex. Since the magnetic structure is deflected while it ascends, the apex location is shifted. Therefore, the transverse plane containing the apex is also changing.  Nevertheless, the $x$-coordinate of the apex of the flux rope remains near the chosen plane during most of the rising phase. We extract the points that correspond to the flux rope in this plane and from these points we trace the magnetic field lines down to the photosphere. These seeds were used to trace the early evolution of the flux rope, as for instance in Fig.~\ref{fig:tempFL}, and for the following analysis of forces and deflection we use the seeds extracted with GUITAR at $x=5$~Mm. 

\begin{figure*}[!ht]
    \centering
    \includegraphics[width=\textwidth]{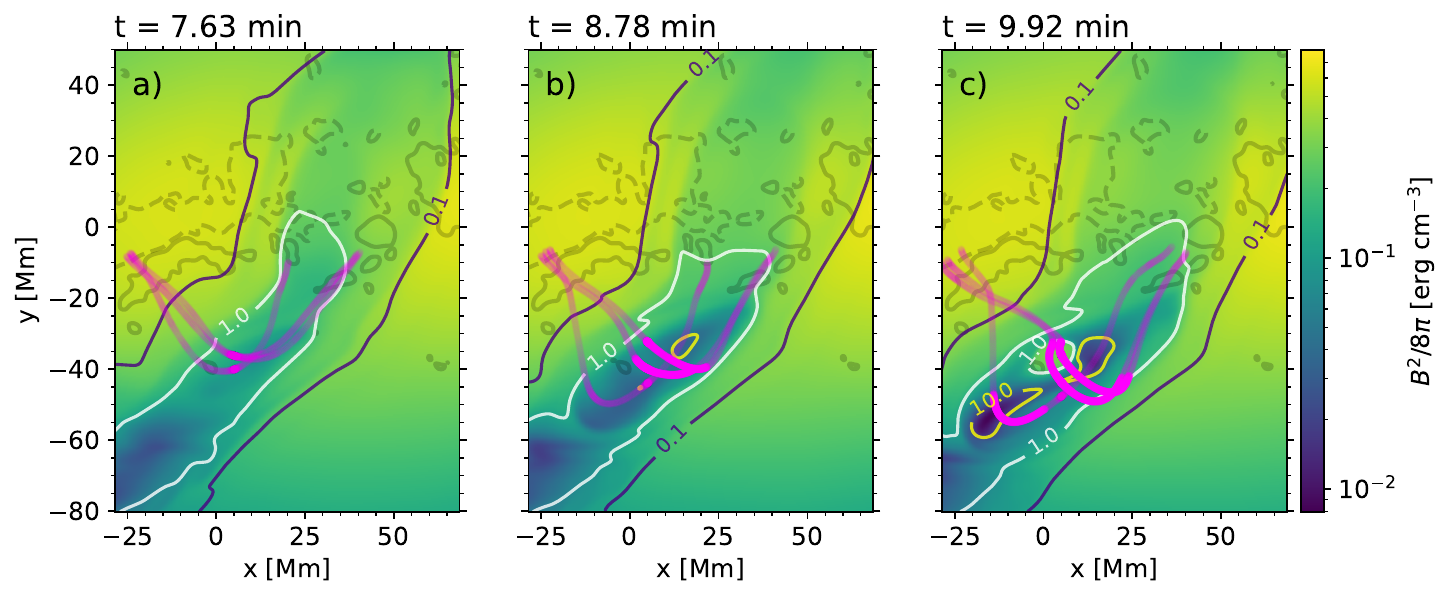}
    \caption{Evolution of the magnetic pressure at $z=60$~Mm for (a) $t=7.63$~min, (b) $8.78$~min and (c) $t=9.92$~min. The background colour shows the variation of $B^2/8\pi$ 
    in blue to yellow colour map, with blue being  the lowest value and yellow the greatest. Different field lines displayed in magenta correspond to the apex, to the left and to the right flanks of the flux rope. The contours show different values of $\beta$, 0.1 (indigo), 1 (white) and 10 (yellow).} 
    \label{fig:B2_zcut}
\end{figure*}

\subsection{Force balance and rising phase}
\label{ss:forces} 

In this section we analyse the forces involved in the upward motion of the flux rope. In the top panels of Fig.~\ref{fig:netFz_zcut} we consider the net vertical force which includes Lorentz force, gas pressure gradient, advection and gravity (all the terms included in the momentum Eq.~\eqref{eq:mom} at $z=24$~Mm, which corresponds to the height where the forces reach their maximum value. The colour scale is shown on the right side from blue (negative) to red (positive) colour  (white corresponds to zero). The panels are taken at a) $t=4.5$~min, b) $t=4.96$~min, c) $t=6.11$~min, d) $t=7.63$~min and e) $t=8.78$~min. The grey contours indicate the positive (solid) and negative (dashed) magnetic field $B_z$ at the photosphere. The green contours represent different values of vertical speed corresponding to 218 (dark green) 
and 436 km s$^{-1}$ (light green). The magenta lines show the field lines belonging to the flux rope, darker segments of these lines are above $z=24$~Mm.  

In the bottom panels of Fig.~\ref{fig:netFz_zcut} we show the 3D magnetic field lines of the flux rope coloured by $v_z$ for the same times as the top panels. As a reference, we also add to this 3D view a semi-transparent plane at $z=24$~Mm together with the contours of $v_z$ in green. 
The rising motion is clearly seen in the 3D field lines of Fig.~\ref{fig:netFz_zcut}, panels h) to j), where the height of the flux rope and the vertical speed increase notably, in contrast to panel f) to h), where the height of the flux rope remains practically constant and there is a gradual increase of the vertical speed. 

Panels~a) to c) of Fig.~\ref{fig:netFz_zcut} show a portion of plasma moving upwards and southwest (y decreases and x increases), indicated by the displacement and enhancement of the net vertical force.  The positive value of this net vertical force comes mostly from the advection term (see Fig.~\ref{fig:Fz_zcut}). This net force is computed with an Eulerian view point, then the local acceleration of the plasma, at a fixed spatial position, is mainly due to the transfer (advection) of the plasma from lower layers.  
As expected, the location of the strong vertical velocity, green contours, is following the same evolution.  
Next the net force decreases, as evidenced from the evolution from panels b) to c).  

Following this upward expansion, the fast rising motion of the flux rope is triggered, indicated in panels c) to e) by the contours of higher $v_z$ and, once again, the enhancement of the net vertical force towards the southeast. At this stage, the increase in the net force is due to the positive total contribution of the gas pressure gradient and the Lorentz force (see Fig.~\ref{fig:Fz_zcut}). The rising motion is accompanied by a deviation of the $x$ apex coordinate of the flux rope, which we call deflection. Before $t=5$~min the apex is moving southwest and after that time it is heading southeast. The deflection will be described in detail in the next section.

\subsection{Deflection}
\label{ss:defl}

During the rising motion, the flux rope is channelled into regions of low magnetic magnetic pressure, causing a deflection of the eruption. This can be observed in Fig.~\ref{fig:B2_zcut}, where we show the magnetic pressure in blue to yellow colour map for $z=60$~Mm, where blue corresponds to lower values and yellow to high values. The different panels represent different times, being a) $t=7.63$~min, b) $8.78$~min and c) $t=9.92$~min. We also display magnetic field lines representing the apex and the left and right flanks of the erupting flux rope in magenta, where the brighter points in it indicate that they are above the horizontal cutting plane ($z=60$~Mm). 
We also include plasma $\beta=p/p_m$ contours, with $p$ corresponding to the gas pressure and $p_m=B^2/8\pi$ the magnetic pressure. The represented values are 0.1 (indigo), 1 (white) and 10 (yellow). The regions where $\beta=10$ corresponds to the flux rope. As the flux rope ascends, magnetic reconnection occurs at the leading edge of the structure, decreasing strongly the magnetic pressure and increasing the plasma pressure, hence, increasing the values of $\beta$. Note that this is a consequence of the evolution of the initial NFFF, since initially there are no regions with $\beta=10$ at this height (see Fig.~\ref{fig:init_atm}, panel b) and it is a highly dynamic feature since the speed of the flux rope is super Alfvénic at this stage (as seen as well with the strong contribution of advection in the evolution of the flux rope in Appendix \ref{s:risingsuppl}).

Figure~\ref{fig:B2_zcut} shows that, as the simulation evolves, the field lines representing the flux rope are deflected towards a region with low magnetic pressure in the $x-y$ plane, denoted by dark blue colours, and crossing the cutting plane where $\beta>1$. We also verify the behaviour of the gas pressure for this height (not shown here), confirming that the gas pressure is increasing but in a manner not as localised as the decrease of the magnetic pressure. This suggests that the deflection is influenced by the decrease of the magnetic pressure, in agreement with previous observational studies \citep{Gui2011, Liewer2015, Sieyra2020}.

\section{Discussion and conclusions}
\label{s:disc_and_concl}

In this work we perform a resistive full-MHD simulation of the flaring active region AR 12241, considering an initial situation composed of a magnetic field obtained from a NFFF extrapolation together with a plane parallel atmosphere ranging from the photosphere to the corona. The extrapolation is based on a vector magnetogram taken by SDO/HMI a few minutes before the onset of the flare. In this simulation, a flux rope forms due to the converging Lorentz force acting on the initial sheared arcade, carrying dense material from the photosphere. In the initial stage, this converging force produces compression and heats locally the plasma above the transition region.
This also results in evaporation of plasma, that fills the flux rope with dense material. Plasma evaporation driven by compression and thermal conduction was demonstrated by \citet{Donne2026} in 2.5D simulations.

As the flux rope rises, as a result of a net upwards force, it is deflected towards regions of low magnetic pressure. This is in agreement with the SDO/AIA observations of this event and with previous observational studies on other deflected eruptive events \citep{Gui2011, Liewer2015, Sieyra2020}. Finally the flux rope leaves the domain after 16 minutes since the initialisation of the simulation with a vertical speed of approximately 350~km s$^{-1}$. To identify the flux rope and characterise its dynamics we use GUITAR \citep{Wagner2024a,Wagner2024b}, which is an algorithm designed for identifying and tracking magnetic structures using the twist as a proxy.

Comparing with other force-free models that minimise the total magnetic energy and assume $\beta\approx0$, the NFFF extrapolation follows an alternative approach minimising the total energy dissipation rate which considers more generalised momentum balance equations supporting a finite plasma pressure \citep[see discussion in][]{Hu2008, Hu2010}. As a result, the magnetic field obtained from this kind of extrapolation presents a non-vanishing Lorentz force in the low atmosphere. On one hand, this allows us to simulate the dynamics of the plasma near the onset of the eruption, having converging and shearing forces from the very beginning. This reduces the computational time of the energy build up and the need of triggering the eruption through the external photospheric driving. On the other hand, since the initial components of the Lorentz force originate from the extrapolation itself, we do not have control on the intensity and direction of the shearing and converging motions. Another limitation is that the parameters $\alpha_i$ that define the NFFF magnetic field ($\mathbf{B}=\mathbf{B}_1+\mathbf{B}_2+\mathbf{B}_3$ where $\nabla \times \mathbf{B}_i=\alpha_i \mathbf{B}_i$ with $i \in (1,2,3)$) are constrained by the size of the horizontal grid (used to normalized $\alpha_i$) since $\alpha_{max}=2\pi/N$, where $N$ is the number of grid points, and the assumption of periodic boundary conditions on the sides (because the calculation of the force free fields includes Fourier transforms). Nevertheless, \cite{Bhattacharyya2007} has demonstrated that arcade structures can be modelled as relaxed states using the minimum dissipation rate principle, hypothesis assumed for NFFF extrapolations. Moreover, there are several recent studies using MHD simulation initiated with the NFFF that successfully explained various transient events in the active regions such as flares, coronal jets, and coronal dimmings \citep{Prasad2018, Prasad2020, Prasad2023, Nayak2019}.

In terms of kinematics, Fig.~\ref{fig:FRap} shows that the vertical speed for the apex of the flux rope is in agreement with the fast phase speed of the observed flux rope shown in Fig.~7~b) from \citet{Joshi2017} (taking into account the observed speed is a projected value on the plane of sky). The slow phase in our model lasts until $\sim 6$~min. Since the model is initiated very close in time to the onset of the flare with an imbalanced magnetic field, the initial compression and twist build up of the flux rope occur faster than in observations ($\sim 18$~min). After the slow phase finishes, around 6~min, the erupting flux rope exhibits a ballistic propagation with a constant acceleration of 424~m\,s$^{-2}$ (1.5 larger than the solar gravity). We can consider that this is the starting of the fast or accelerated phase. However, as also stated by  \citet{Jiang2024}, it is difficult to determine precisely the beginning of the eruption phase.
In the non-ideal models from the literature, the acceleration of the eruptive flux rope is related to magnetic reconnection, which causes the impulsive acceleration. In \citet{Inoue2025} they show how a small flux emergence reconnects with the ambient field. Comparing simulations, one obtained from an evolved data-driven model and the other a data-constrained including flux emergence, both initialised with NLFFF extrapolations and plasma $\beta=0$, they obtained that at the initial stage both simulations are dynamically similar. However, just after the triggering of the acceleration phase, the vertical velocities start to diverge, with the data-constrained simulation showing larger acceleration than the data-driven one.

Another point to consider is the atmospheric profile that was chosen here, and how it influences the dynamics, density and temperature of the flux rope. To address this point we performed another simulation with the same magnetic field but with a smoother and higher transition region (still using a plane-parallel stratified initial condition). The simulation is not shown here for the sake of conciseness, but it shows qualitatively the same behaviour as the simulation presented in this work. Still, some differences are notable when considering this wider transition region. First, the dynamics are a bit altered, and the eruption is delayed for more than 6 minutes. Similar results were shown by \citet{Rice2025} with 2.5 MHD simulations considering different temperature profiles. In this other simulation the rising phase starts later than in the simulation presented in this work, because heavier material takes more time to lift up, and hence, the flux rope leaves the domain later with a lower speed of 250~km s$^{-1}$. Second, concerning the thermodynamical structure of the flux rope, changing the profile of the transition region modifies the density, and temperature of the flux rope. We found for the higher transition region case, the flux rope is almost twice as dense and cold with respect to the atmosphere profile analysed in this work, and this has an impact on the emission. Therefore, aiming to reproduce observational features, it is of upmost importance to carefully consider a realistic atmosphere. In a forthcoming paper, we aim to explore quantitatively the synthetic emission obtained from the simulations, not included here because the focus of the article is on presenting the model and describing the dynamics. 

And lastly, our compressible full MHD simulation results are comparable with the outcomes from the incompressible and isothermal simulation of \citet{Prasad2023} (see discussion in Introduction) in the fact that the triggering of the eruption is the converging Lorentz force at the lower atmosphere exerted on a sheared arcade that develops later into a flux rope. However, they pointed out that in their simulations the rising motion of the flux rope ceased towards the end whereas in our case the flux rope leaves the domain at a constant acceleration. We also notice that gravity can slow down the escaping FR, but in this case it acts too slowly to affect its speed. The two simulations start from the same initial extrapolation, which strikingly shows that our use of fully compressible MHD affects the predicted characteristics of the escaping flux rope. 

As future prospects, we intend to include flux emergence in our simulation framework \citep[as for example in][]{Inoue2025, Moreno-Insertis2025}.
We are also working towards estimating non-thermal emissions from the inclusion of test-particles in the simulation.

\begin{acknowledgements}
This work has been supported by the French Agence Nationale de la Recherche (ANR) project STORMGENESIS \#ANR-22-CE31-0013-01, the European Research Council through the Synergy Grant number 810218 (``The Whole Sun'', ERC-2018-SyG), the Centre National d’Etudes Spatiales (CNES) Solar Orbiter project, the Institut National des Sciences de l’Univers (INSU) via the Action Thématique Soleil-Terre (ATST), and Programme for Supercomputing, and by computing HPC and storage resources by GENCI thanks to the grants 2023-A0140410133, 2024-A0160410133 and 2025-A0180410133. AW acknowledges the Finnish Centre of Excellence in Research of Sustainable
Space (Research Council of Finland grant numbers 352850). AW also acknowledges the Space Weather Awareness Training Network (SWATNet) which has received funding from the European Union’s Horizon 2020 research and innovation programme under the Marie Skłodowska-Curie Innovative Training Networks, Grant Agreement No 955620.
RJ acknowledges research support from the Research Council of Norway, project number 325491, through its Centres of Excellence scheme, project number 262622.
\end{acknowledgements}

%
\bibliographystyle{aa}
\bibliography{biblio}

\begin{appendix}
\section{Numerical setup}
\label{s:setup}
For the numerical framework developed in this work we use the PLUTO code \citep[][\url{https://plutocode.ph.unito.it/}]{Mignone2007} which is a modular freely-distributed software that solves mixed hyperbolic/parabolic systems of partial differential equations (conservation laws) targeting high Mach number flows in astrophysical plasma dynamics. In the following sections we describe the technical details, such as the numerical schemes, the solved equations, the initial and boundary conditions.

\subsection{MHD equations}
\label{ss:eq}
The adimensional compressible resistive MHD equations solved by PLUTO are:

\begin{eqnarray}
\frac{\partial \rho}{\partial t} &+& \nabla \cdot (\rho \mathbf{v}) = 0 \label{eq:cont}\\
\frac{\partial \rho \mathbf{v}}{\partial t} &+& \nabla \cdot [\rho \mathbf{v} \mathbf{v} - \mathbf{B B} + \mathrm{I} (p + \frac{\mathbf{B}^2}{2})] = \rho \mathbf{g} \label{eq:mom}\\
\frac{\partial \mathbf{B}}{\partial t} &+& \nabla \times (-\mathbf{v} \times \mathbf{B}) = - \nabla \times \eta (\mathbf{\nabla \times \mathbf{B}}) \label{eq:ind}\\
\frac{\partial E_t}{\partial t} &+& \nabla \cdot [(E_t + p_t) \mathbf{v} - \mathbf{B} (\mathbf{v} \cdot \mathbf{B}) + \eta (\nabla \times \mathbf{B}) \times \mathbf{B} ] \nonumber \\ &=&\rho \mathbf{v} \cdot \mathbf{g} - \nabla \cdot \mathbf{F_c} - n^2 \Lambda (T) \, ,\label{eq:energy}
\end{eqnarray}
where $\rho$ is the plasma density, $p$ is the gas pressure, $\mathbf{v}$ and $\mathbf{B}$ are the velocity and magnetic fields. $E_t = \frac{\rho v^2}{2} + \rho e + \frac{B^2}{2}$  is the total energy density and $p_t = p + \frac{B^2}{2}$ the total pressure (gas pressure and magnetic pressure), $e$ is the internal 
energy per unit mass. In these units, a factor $1/\sqrt{4\pi}$ is included in the definition of the magnetic field ${\bf B}$. The parameters $\eta=1 \times 10^{11}$~cm$^2$ s$^{-1}$ and $\mathbf{g}=-\frac{G M_{\odot}}{R_{\odot}^2}\,{\bf e}_z=-27394\,{\bf e}_z$~cm s$^{-2}$ are the magnetic diffusivity and gravity at the solar surface, respectively. Both quantities are uniform and constant for the entire domain and the duration of the simulation.

In Eq.~\eqref{eq:energy}, $\mathbf{F}_c$ is the thermal conduction flux, which varies between the classical (Spitzer-H\"arm) and saturated thermal conduction regimes $\mathbf{F}_{\rm class}$ and $F_{\rm sat}$, respectively, and is given by:

\begin{eqnarray}
\mathbf{F}_c &=& \frac{F_{\rm sat}}{F_{\rm sat} + |\mathbf{F}_{\rm class}|} \mathbf{F}_{\rm class} \label{eq:fc} \\
\mathbf{F}_{\rm class} &=& -\kappa_{\parallel}( {\bf b} \cdot \nabla T)  {\bf b} \label{eq:fclass} \\
F_{\rm sat} &=& 5\, \phi\, \frac{p^{3/2}}{\rho} \, ,\label{eq:fsat}   
\end{eqnarray}
where ${\bf b}= \frac{{\bf B}}{\rm{B}}$ is a unitary vector parallel to the magnetic field, $\kappa_{\parallel}=5.6 \times 10^{-7} T^{5/2}$~erg s$^{-1}$ K$^{-1}$ cm$^{-1}$, $\phi = 0.3$. In our study, we only consider thermal conduction along the magnetic field. $T$ corresponds to the plasma temperature and is given in Kelvin (K).

In Eq.~\eqref{eq:energy}, we also have the cooling term that depends on $n$, the density number, and $\Lambda$, the radiative loss function \citep{Athay1986} in the following form:
\begin{eqnarray}
    \Lambda = 10^{-22} &\{&0.4 \exp[-30 (\log T-4.6)^2] \nonumber \\ &+&4 \exp[-20(\log T-4.9)^2] \nonumber \\
            &+&4.5\exp[-16(\log T-5.35)^2]\nonumber \\ &+&2\exp[-4(\log T-6.1)^2]\} \, , 
    \label{eq:rad}
\end{eqnarray} 
where $T$ is in K and $\Lambda$ is in erg s$^{-1}$. This formulation was used successfully in a global stellar wind model by \citet{Reville2020}. It is not considered to be very precise at low temperature (typically below $10^4$), but reproduces correctly the overall trend of the cooling rate as a function of temperature. It was used here for the sake of its simplicity, more precise parametrizations in particular for the chromosphere will be used in future works (e.g. using CHIANTI, see \citealt{Reville2024}).

From Eq.~\eqref{eq:energy}, we can derive an equation for the evolution of the internal energy $E_{\rm{int}}=\rho e=p/(\Gamma -1)$, where $\Gamma$ is the ratio of the specific heats $c_p/c_v$, as follows:

\begin{eqnarray}
    \frac{\partial E_{\rm{int}}}{\partial t}&=&-\nabla \cdot \mathbf{F}_c
    -\nabla \cdot \mathbf{F}_{\rm{int}}+\mathbf{v} \cdot \nabla p  
    + \eta |\nabla \times \mathbf{B}|^2 
    - n^2 \Lambda(T) \nonumber \\
    &=&-\nabla \cdot \mathbf{F}_c 
     - \frac{\Gamma}{\Gamma -1}\, p\,\nabla \cdot \mathbf{v} 
     - \mathbf{v} \cdot \nabla \frac{p}{\Gamma -1} 
     + \eta |\nabla \times \mathbf{B}|^2 \nonumber \\&& - n^2 \Lambda(T) \, ,
    \label{eq:int_energy}    
\end{eqnarray}
where $\mathbf{F}_{\rm{int}}=(p + E_{\rm{int}})\mathbf{v}$ is the enthalpy flux. The second and third terms on the right-hand side correspond to the compression and advection components analysed in Sect.~\ref{ss:Compression}.

We solve the system of Eq.~\eqref{eq:cont}--\eqref{eq:energy} in 3D cartesian coordinates using a 2\textsuperscript{nd} order Runge-Kutta for advancing the solution in time, the Harten, Lax, Van Leer (HLL) approximate Riemann Solver and linear reconstruction. To preserve $\nabla \cdot \mathbf{B}=0$, we consider the constrained transport method \citep{Londrillo2004,Gardiner2005}, where two sets of magnetic fields (face-centered and cell-centered) are used to calculate the electric field. We close the system with using the perfect gas law with constant specific heat as an equation of state.

\subsection{Grid}
\label{ss:grid}
The computational domain spans
$\mp138$~Mm in $x$ with a total of 884 cells, from $-124$ to $62$~Mm in $y$ with 554 cells and from $0$ to $186$~Mm in the vertical direction $z$ with 1024 cells. The grid spacing is uniform in $x$ with a resolution of $\Delta x=312$~km. In $y$ and $z$ directions the grid is mixed, having a uniform and stretched spacing. The horizontal $y$-coordinate is uniform, with $\Delta y_{\rm{min}}=312$~km, in the area where the magnetic field is strong, $y=\mp62$~Mm, comprising 398 cells, and stretched, with $\Delta y_{\rm{max}}=496$~km, in the region where the magnetic field is weak, from $-62$ to $-124$~Mm with 156 cells. For the vertical coordinate $z$, the spacing is uniform with $\Delta z_{\rm{min}}=104$~km in the lower atmosphere and beyond the transition region, from 0 to $10$~Mm, comprising 96 cells, and is stretched in the corona, spanning from $10$~Mm to $186$~Mm with 928 cells resulting in $\Delta z_{\rm{max}}=312$~km. The stretched grid in PLUTO is defined as $r \, (1-r^N)/(1-r) = (x_R-x_L)/(\Delta x)$, where $r$ is the stretching ratio, $N$ is the number of points in the stretched grid, $\Delta x$ is the uniform grid resolution and $x_L$ and $x_R$ are the leftmost and rightmost points of the stretched grid. 

\subsection{Initial conditions}
\label{ss:ic}

\begin{figure*}[!ht]
    \centering
    \includegraphics[width=\textwidth]{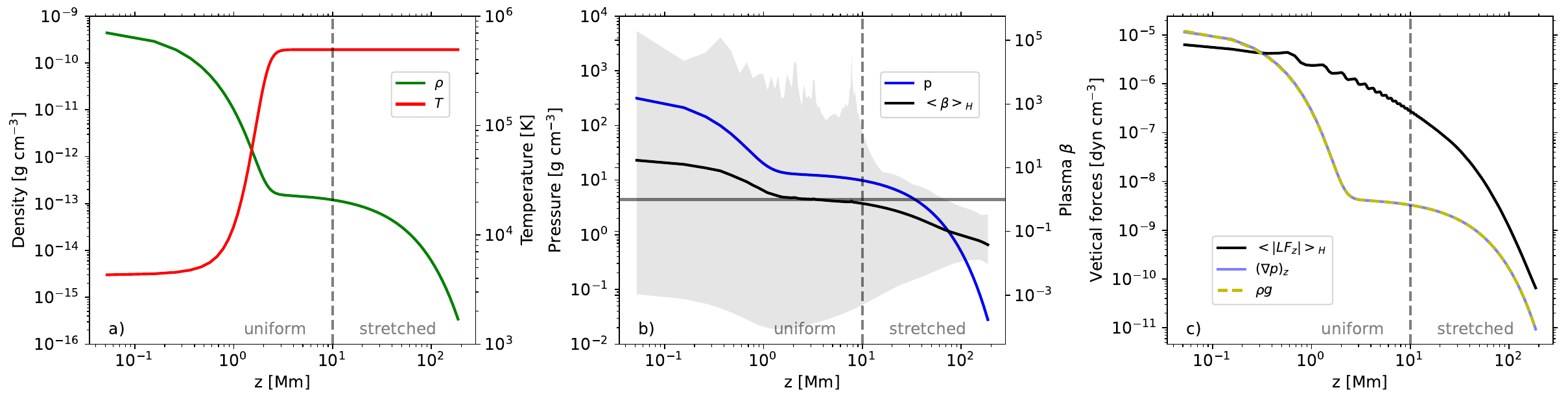}
    \caption{Vertical initial profiles. 
    a) Density ($\rho$, green solid line) and temperature ($T$, red solid line) as function of height. 
    b) Gas pressure ($p$, blue solid line) and plasma $\beta$ ($<\beta>_H$, black solid line). The grey area covers the minimum and the maximum values of $\beta$ for each height. The horizontal grey line indicates $\beta=1$.  
    c) Initial forces: horizontal average of the vertical Lorentz force ($<LF_z>_H$, black solid line), gas pressure gradient ($(\nabla p)_z$, blue solid line) and gravity ($\rho g$, dashed yellow line). The vertical dashed grey line indicates the division between uniform and stretched grid in $z$.}
    \label{fig:init_atm}
\end{figure*}

\begin{figure*}[!ht]
    \centering
    \includegraphics[trim={0cm 2cm 0cm 6cm},clip, width=\linewidth]{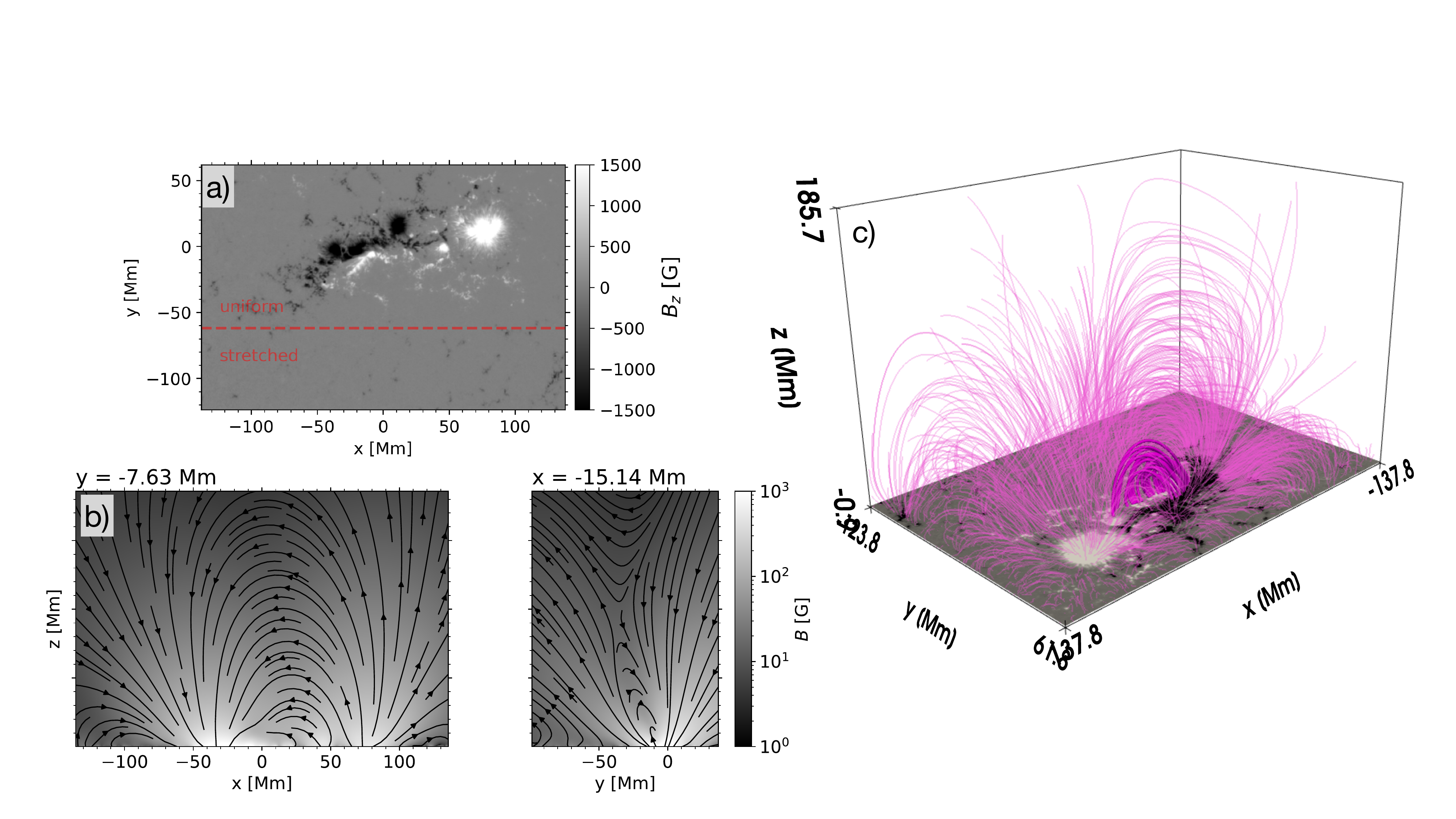}
    \caption{Initial magnetic field. a) Initial $B_z$ at $z=0.05$~Mm in grey scale, positive (negative) values are in white (black). The dashed red line indicates the division between the uniform and stretched grid in $y$. b) Cuts on vertical planes at  $y=-7.63$~Mm (left) and $x=-15.14$~Mm (right). The magnitude of the magnetic field is shown in grey scale and the black lines correspond the streamlines of the projected magnetic field vector on each plane. c) 3D visualization of $\mathbf{B}$ (thin magenta lines), with the flux rope field lines highlighted (thick magenta lines).}
    \label{fig:B_init}
\end{figure*}

We initialise the simulation using a plane-parallel stratified atmosphere together with a NFFF magnetic field extrapolation. For the magnetic field we applied a divergence cleaning method described by \citet{Valori2013}, in which we correct the initial magnetic field $\mathbf{B}$ as 
$\mathbf{B}_s=\mathbf{B}+\hat{k}\int^{z_2}_z (\nabla \cdot \mathbf{B}) dz'$. 
This ensures $\nabla \cdot \mathbf{B}_s=0$ to machine accuracy.  Note that this correction only changes $B_z$, leaving the vertical component of the current density $(\nabla \times \mathbf{B})_z$ unchanged. 

Regarding the thermodynamical variables, we choose the density as a combination of two exponential functions:  
   $$\rho = \rho_1\exp\left[\frac{-(z - z_{ch})}{z_{c1}}\right] 
          + \rho_2\exp\left[\frac{-(z - z_{ch})}{z_{c2}}\right] \, .$$ 
The values chosen for this simulation were $\rho_1=4\times 10^{-10}$~g cm${-3}$, $\rho_2=1.67\times 10^{-13}$~g cm${-3}$, $z_{ch}=0.075$~Mm, $z_{c1}=0.25$~Mm and $z_{c2}=30$~Mm. This gives density values ranging from $3\times 10^{-16}$ to $3\times 10^{-10}$~g cm$^{-3}$, and, from $3\times 10^{-2}$ to 300~dyn cm$^{-2}$ for the gas pressure. 

From the hydrostatic equilibrium equation we obtain the profile for the gas pressure. The temperature $T$ is defined with the thermal equation of state $p=k_B\,\rho\,T/m_u\,\mu$, where $k_B$ is the Boltzmann constant, $m_u$ is the atomic mass unit and $\mu$ the mean molecular weight. We assume that the plasma is fully ionized ($\mu=0.5$). We obtain that the temperature varies from $4 \times 10^3$ to $5 \times 10^5$~K. Fig.~\ref{fig:init_atm} summarises different initial variables as a function of height. Panel a) displays the temperature (red solid line) and the density (green solid line). Panel b) displays the gas pressure (blue) together with the horizontal average of the plasma $\beta$ (black). The horizontal grey line indicates $\beta=1$. The grey area around $<\beta>_H$ spans from the minimum value of plasma-$\beta$ (bottom limit) to the maximum value (upper limit) at each height. The amplitude of these values shows the non-uniformity of this parameter, especially below the transition region (located at $z \approx 1-2$ Mm), being $<\beta>_H\,<1$ above 7~Mm and entirely lower than one above 70~Mm. 
The large values of $\beta$ below the corona ($z<3$~Mm) correspond to small localised regions
in the quiet Sun, in opposite directions where the eruption will be developed. Panel c) exhibits the variation of the horizontal average of the Lorentz force norm with height (black) together with the gas pressure gradient in the vertical direction (blue)  and the gravity force (dashed yellow). The vertical dashed grey lines in all the panels show the limit where we have used a uniform grid (to the left of this line) and a stretched grid (to the right) in $z$. 

Figure~\ref{fig:B_init} shows the initial magnetic field obtained from the NFFF extrapolation. 
Panel a) shows the vertical component $B_z$ at the $z=0.05$~Mm and the dashed-red line delimits the region where the grid is uniform in $y$ (above the red line) and stretched (below it). Note that the active region is not centred with the $y$-coordinate of the simulation box. The reason of this choice is that, from observations, the eruption is ejected towards the south (negative $y$ in our simulation box), thus we only extended the domain in this direction to maximize numerical efficiency. 
Panel b) shows the magnitude of the magnetic field in grey scale in two perpendicular planes 
at around $y=-7$~Mm and $x=-15$~Mm. The black streamlines represent the projected magnetic field vector on each plane. Panel c) shows a 3D view of the magnetic field (magenta thin lines) together with the field lines that will form the flux rope (thick magenta lines).

\subsection{Boundary conditions}
\label{ss:bc}
For the bottom boundary, we use the line-tied conditions as expressed by \citet{Aulanier2005}, in which all the components of the velocity are set to zero for the entire duration of the simulation, meaning that $v_x^{i,j,K_{\rm{BEG}}}=v_y^{i,j,K_{\rm{BEG}}}=v_z^{i,j,K_{\rm{BEG}}}=0$, where $K_{\rm{BEG}}$ corresponds to the first index of the physical domain in the vertical coordinate $z$, $i=I_{\rm{BEG}},...,I_{\rm{END}}$ and $j=J_{\rm{BEG}},...,J_{\rm{END}}$ corresponds to the horizontal coordinates $x$ and $y$, respectively. For the ghost cells we define symmetry for $\rho$ and antisymmetry for $v_z$ that proved to be efficient at enforcing line-tying (e.g. \citealt{Aulanier2005}) as:
   $$\rho^{i,j,K_{\rm{BEG}}-k}=\rho^{i,j,K_{\rm{BEG}}+k}$$
   $$v_z^{i,j,K_{\rm{BEG}}-k} =-v_z^{i,j,K_{\rm{BEG}}+k} \, ,$$
where $k=1,2$. The third index on the left hand side spans the two ghost cells and the one in the right hand side extends over the physical domain.  
For all the remaining variables, we define second derivative centred at $K_{\rm{BEG}}$ to be zero (linear extrapolation), resulting in:
   $$f^{i,j,K_{\rm{BEG}}-k} = 2 f^{i,j,K_{\rm{BEG}}}-f^{i,j,K_{\rm{BEG}}+k} \, ,$$
where $f=\{p,v_x,v_y,B_x,B_y,B_z\}$.

For the side boundaries, we set zero-gradient conditions for $\rho,p$ and $\mathbf{v}$, fixing to zero the normal component of velocity in case it goes inwards, to avoid having inflows. For the magnetic field, we consider linear extrapolation  with centred derivative at $I_{\rm{BEG}}$ and $I_{\rm{END}}$ for the planes $yz$ as follows:
   $$B_{x,y,z}^{I_{\rm{BEG}}-i,j,k}=2 B_{x,y,z}^{I_{\rm{BEG}},j,k}-B_{x,y,z}^{I_{\rm{BEG}}+i,j,k}$$   
$$B_{x,y,z}^{I_{\rm{END}}+i,j,k}=2 B_{x,y,z}^{I_{\rm{END}},j,k}-B_{x,y,z}^{I_{\rm{END}}-i,j,k} \, ,$$
where $i=1,2$. 
For the $xz$ planes, we use:
$$B_{x,y,z}^{i,J_{\rm{BEG}}-j,k}=2 B_{x,y,z}^{i,J_{\rm{BEG}},k}-B_{x,y,z}^{i,J_{\rm{BEG}}+j,k}$$
$$B_{x,y,z}^{i,J_{\rm{END}}+j,k}=B_{x,y,z}^{i,J_{\rm{END}},k} \, ,$$
where $j=1,2$. The asymmetry in this condition is due to the end boundary in the $y$-direction is still close to the centre of the active region, as it was not extended as much as in the other $y$-direction in order to minimise the numerical cost of the simulations. When using a linear extrapolation for the magnetic field there, we found that spurious inflows could form and affect this part of the simulation domain (while not affecting the eruption under study here itself, as it happens on the other side of the active region). Still, to avoid these spurious boundary effect we found that using a zero-gradient boundary condition on this side of the box prevented the simulation to exhibit any discernable effect from this lateral boundary. 
We keep these conditions for the magnetic field on the sides until the shock caused by the initial instability has left the domain, which is around $t=4.20$~min. After that time, we use the zero-gradient for $\mathbf{B}$ for all the side boundaries.

For the top boundary, we apply zero-gradient for $\rho,p,v_x, v_y$ and $v_z$, setting $v_z=0$ if it is pointing inward, and linear extrapolation for the three components of magnetic field.

\section{GUITAR}
\label{s:guitar}

GUITAR is a graphical user interface, designed for extracting the flux rope field line source points from 2D flux rope proxy maps \citep{Wagner2024b}. Commonly-used thresholds like the twist number $\mathcal{T}_w$ or the winding number $\mathcal{T}_g$ can serve as proxy for the location of magnetic flux ropes \citep[see e.g. the appendix in][for an extensive discussion on these twist metrics]{Liu2016}. However, often this is not a sufficient criterion as multiple twisted structures in simulation data may be present. To improve the results of a thresholding procedure, GUITAR incorporates mathematical morphology algorithms namely dilatation and erosion. These are built upon comparing image elements with a pre-defined shape with variable size called the structuring element (SE). 

We apply these algorithms to thresholded $\mathcal{T}_w$ maps. More precisely we use a binary formalism of them. Let X be a binary image. The selected region, called C, corresponds to values 1 within X (which are not necessarily connected), and SE is the structuring element with centre point $s$. Then the dilation of X (X $\oplus$ SE) is the union of all points $x$ $\in$ X that the SE can reach, with SE centred at any point $c$ $\in$ C. Thus, the dilation operation adds values 1 at the edges of C, depending on the shape and size of SE. On the other hand, the erosion of X (X $\ominus$ SE), is the union of all centre points $s$ of the SE for which SE is fully contained in C. Thus, the erosion operation removes values 1 at the edges of C, depending on the shape and size of SE. Note these algorithms are (in most cases) not inverse to each other and combining them in particular ways results in very useful new procedures. 

Most notably, for the applications in GUITAR, the opening algorithm corresponds to the dilation of an erosion: (X $\ominus$ SE) $\oplus$ SE. It is typically used to remove noise or unwanted (sub-)features.  This is particularly useful for post-processing the thresholding of $\mathcal{T}_w$ (or other flux rope proxy) maps. A circle is selected for SE in the case of GUITAR, to avoid the flux rope cross section to contain sharp edges in regions  where processing is applied. 
Once the user is satisfied with the results, the desired feature can be tracked through time.  Lastly, the source points can be saved either via uniform or random sampling within the extracted binary shapes. Finally, it should be noted that GUITAR is also designed to customize and save plots and animations, export intermediate steps during the processing and do a basic 2D analysis of the time series of the extracted binary shapes, for example, the flux rope cross-section \citep{Wagner2024b}.

\section{Magnetic reconnection}
\label{s:reconnection}

\begin{figure*}
    \centering
    \includegraphics[trim={0cm 0cm 0cm 14cm},clip,width=\textwidth]{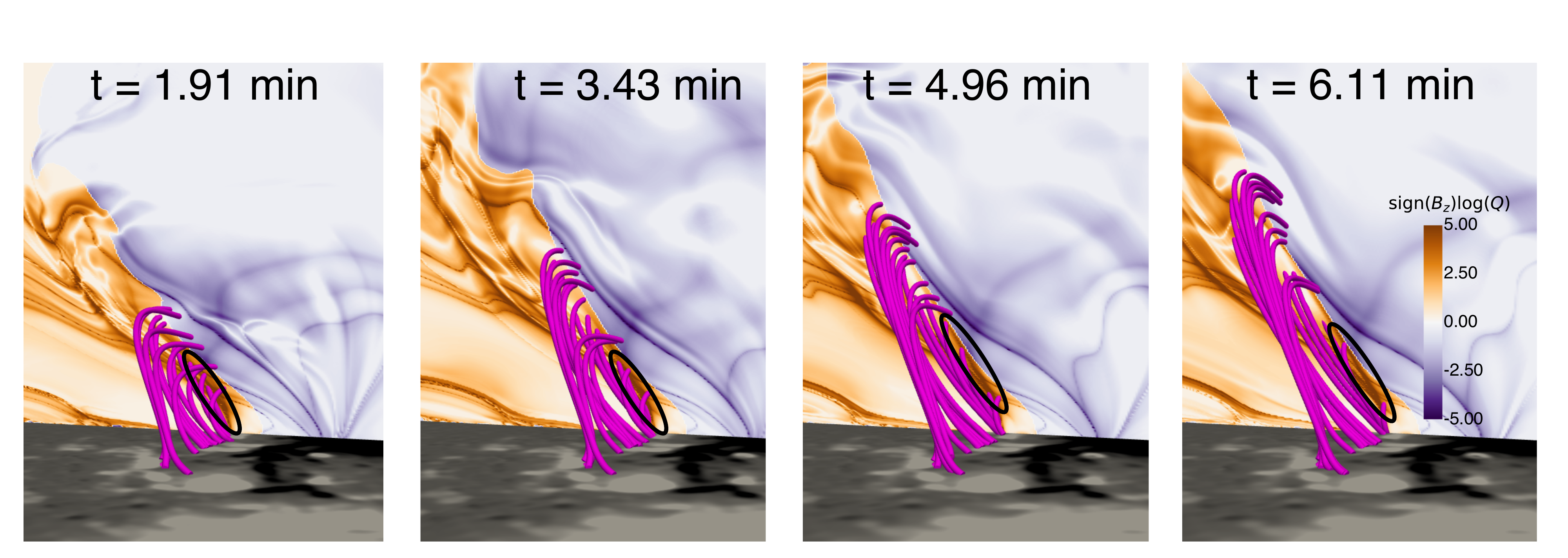}
    \caption{Early evolution of the squashing factor $Q$. The panels display ${\rm{sign}}(B_z){\rm{log}}(Q)$ for a cutting plane at $x=5$~Mm in purple (negative $B_z$) to orange (positive $B_z$) colours for $t=1.91$~min, $t=3.43$~min, $t=4.96$~min, and $t=6.11$~min. The magenta lines correspond to the flux rope field lines. The black ellipses indicate regions of intersection of the field lines with high $Q$ values areas.}
    \label{fig:sq}
\end{figure*}

To inspect the presence of magnetic reconnection in the simulation, we compute the squashing factor $Q$, which measures the deformation of the field line mapping between two surfaces. The regions where the gradients in the mapping are large are referred as quasi-separatrix layers (QSLs), and are a continuous extension of separators and separatrix surfaces. These QSLs are potential sites for the formation of strong electric currents, and therefore, possible regions where magnetic reconnection can occur. For the calculations of $Q$ we use the package K-QSL \url{https://github.com/Kai-E-Yang/QSL/tree/master} which follows the method described in \cite{Scott2017} and \citet{Tassev2017}. The results of ${\rm{sign}}(B_z){\rm{log}}(Q)$ for a cutting plane at $x=5$~Mm and different times are shown in Fig.~\ref{fig:sq}. High values of $Q$ are denoted in orange (for $B_z>0$) and dark purple (for $B_z<0$). The field lines corresponding to the flux rope are displayed in magenta. The black ellipses highlight the regions where the flux rope field lines cross through areas of high $Q$, suggesting that these field lines are reconnecting. The times considered in this figure correspond to a stage before the triggering of the rising phase ($t=1.91$~min), at the beginning and during it ($t=3.43$~min and $t=4.96$), and for a later time where there is a clear increase of the vertical velocity of the flux rope ($t=6.11$~min, see panel c) from Fig.~\ref{fig:netFz_zcut} as a reference).

\section{Force Analysis}
\label{s:risingsuppl}

\begin{figure*}[!ht]
    \includegraphics[width=\textwidth]{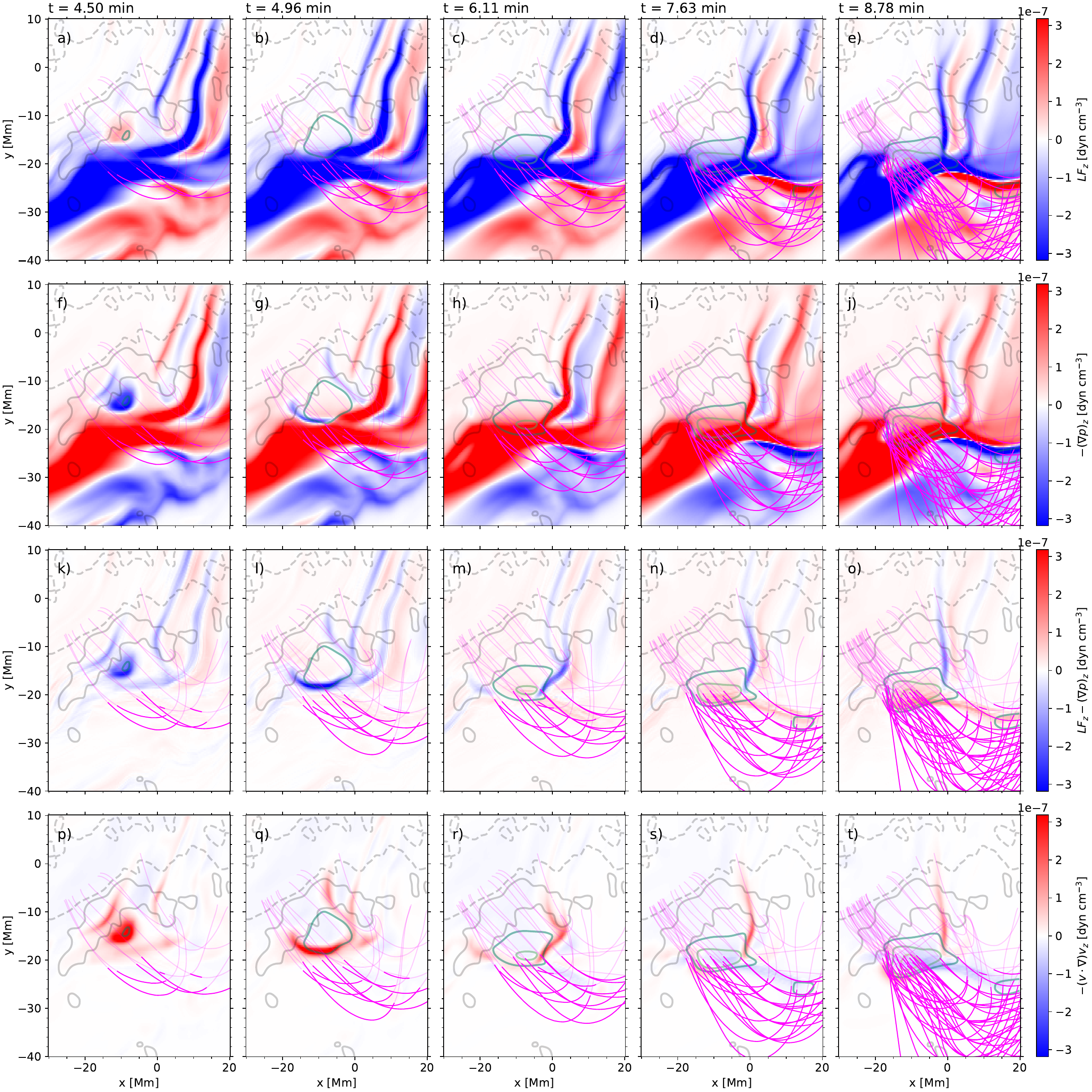}
    \caption{Evolution of the vertical forces  (positive values are in red and negative values in blue) at $z=24$~Mm for t=4.5 min, t=4.96 min, t=6.11 min, t=7.63 min and t=8.78 min. Panels a) to e) show the vertical component of the Lorentz force, f) to j) display the negative gas pressure gradient, k) to o) exhibit the net contribution of the Lorentz force and gas pressure gradient and panels p) to t) the advective component of the force. The magenta points correspond to the flux rope field lines, being darker those above the cutting plane. grey contours indicate positive (solid) and negative (dashed) values of $B_z$ at $z=0.05$~Mm and the green contours exhibit different values of $v_z$ between 218 (dark green) and 436 km s$^{-1}$ (light green).}
    \label{fig:Fz_zcut}
\end{figure*}

To complement the analysis discussed in Sect.~\ref{s:Rising}, we show in Fig.~\ref{fig:Fz_zcut} the forces that contribute to the net force displayed in Fig.~\ref{fig:netFz_zcut}. We exclude gravity in the figure because it is almost negligible compared to the other forces. These forces correspond to each term of the divergence operator (which signs are as if they were on the right hand side of the equation) and the source term of Eq.~\eqref{eq:mom}. The columns indicate the simulation time (which is the same as in Fig.~\ref{fig:netFz_zcut}) and the rows represent the different forces in the vertical direction $z$ with in blue (negative) to red (positive values). Panels a) to e) show the vertical component of the Lorentz force $\rm{LF}_z$, f) to j) exhibit the negative vertical component of the gas pressure gradient, the net contribution between these two is shown in panels k) to o) and in panels p) to t) the contribution of the advection term. The grey contours delimit the positive (solid line) and negative (dashed) values of $B_z$ on the bottom plane $z=0.05$~Mm and the green contours indicate values of $v_z$ on the cutting plane $z=24$~Mm, from 218 (dark green) to 436 km s$^{-1}$ (light green). The magenta lines represent flux rope magnetic field lines, the darker (lighter) segments of the lines are above (below) the cutting plane. 

The contribution of the Lorentz force is always negative in the region where we have higher $v_z$ (green contour), which is near the region where the magnetic field lines are emerging the cutting plane, except for $t=4.50$~min. The opposite happens with the pressure gradient force $-(\nabla p)_z$, where we find positive contribution in the high speed regions. These two forces mostly cancel each other as shown in panels k) to  to o). In the region of higher vertical velocity (within and nearby the dark green contour), the net contribution is first negative (panels k) and l)), then positive (panels n) and o)). 
This indicates that the contribution of the gas pressure gradient is slightly larger than that of the Lorentz force.  

Regarding the advection, panels p) to t), its contribution to the source term of Eq.~\eqref{eq:mom} is positive for the gradual expansion phase, from $t=4.50$ to $t=4.96$~min, and negative for the rising phase, from $t=6.11$ to $t=8.78$~min.  Since the total force is positive (within and nearby the dark green contour), as shown in Fig.~\ref{fig:netFz_zcut}, the Eulerian increase of velocity is first mainly due to the convective transport from below, and then to the pressure gradient force.

\clearpage

\end{appendix}
\end{document}